\documentclass[prx,aps,superscriptaddress,10pt,floatfix,twocolumn,amsmath,amssymb]{revtex4-1}

\usepackage[utf8]{inputenc}
\usepackage[]{graphicx} 
\usepackage{layouts}
\usepackage[dvipsnames]{xcolor}
\usepackage{hyperref}
\usepackage{braket}
\usepackage{ragged2e}

\DeclareMathOperator{\Tr}{Tr}

\begin{document}

\title{Single-Phonon Addition and Subtraction to a Mechanical Thermal State}

\author{Georg~Enzian}
\affiliation{QOLS, Blackett Laboratory, Imperial College London, United Kingdom}
\affiliation{Clarendon Laboratory, Department of Physics, University of Oxford, United Kingdom}
\affiliation{Niels Bohr Institute, University of Copenhagen, Denmark}

\author{John~J.~Price}
\affiliation{QOLS, Blackett Laboratory, Imperial College London, United Kingdom}
\affiliation{Clarendon Laboratory, Department of Physics, University of Oxford, United Kingdom}
\affiliation{Centre for Photonics and Photonic Materials, Department of Physics, University of Bath, United Kingdom}

\author{Lars~Freisem}
\affiliation{QOLS, Blackett Laboratory, Imperial College London, United Kingdom}
\affiliation{Clarendon Laboratory, Department of Physics, University of Oxford, United Kingdom}

\author{Joshua~Nunn}
\affiliation{Centre for Photonics and Photonic Materials, Department of Physics, University of Bath, United Kingdom}

\author{Jiri~Janousek}
\author{Ben~C.~Buchler}
\author{Ping~Koy~Lam}
\affiliation{Centre for Quantum Computation and Communication Technology, Research School of Physics and Engineering, Australian National University, Canberra, Australia}

\author{Michael~R.~Vanner}
\affiliation{QOLS, Blackett Laboratory, Imperial College London, United Kingdom}
\affiliation{Clarendon Laboratory, Department of Physics, University of Oxford, United Kingdom}

\begin{abstract}
Adding or subtracting a single quantum of excitation to a thermal state of a bosonic system has the counter-intuitive effect of approximately doubling its mean occupation. We perform the first experimental demonstration of this effect outside optics by implementing single-phonon addition and subtraction to a thermal state of a mechanical oscillator via Brillouin optomechanics in an optical whispering-gallery microresonator. Using a detection scheme that combines single-photon counting and optical heterodyne detection, we observe this doubling of the mechanical thermal fluctuations to a high precision. The capabilities of this joint click-dyne detection scheme adds a significant new dimension for optomechanical quantum science and applications.
\end{abstract}

\maketitle

\textit{Introduction.}---Performing single-quantum-level operations to bosonic quantum systems provides a rich avenue for quantum-state engineering, quantum-information and communication applications, as well as exploring the foundations of physics. Prominent examples of quantum-state engineering using such operations include the generation of non-classical states of motion of trapped ions~\cite{Leibfried1996}, microwave fields inside superconducting resonators~\cite{Deleglise2008, Hofheinz2009}, and high-frequency phonons coupled to superconducting qubits~\cite{Chu2018, Satzsinger2018}. This type of control is also central to many quantum technologies such as quantum communications with quantum repeaters~\cite{DLCZ2001, Sangouard2011}, on-demand single-photon preparation~\cite{Laurat2006, Specht2011}, and continuous-variable entanglement distillation~\cite{Ourjoumtsev2007}. Moreover, these operations allow for studies of non-classicality~\cite{Zavatta2004, Paternostro2011}, and exploring the interface between quantum information and quantum thermodynamics~\cite{Vidrighin2016}.

A practical and powerful way to achieve single-quantum addition or subtraction is to use an interaction with light followed by single-photon detection. This approach has been applied in optics to create `kitten' states by single-photon subtraction from squeezed vacuum~\cite{Ourjoumtsev2006, Neergaard2006}, and to explore the properties of single-photon-added coherent states~\cite{Zavatta2004}. Single-quantum addition via single-photon detection has also been recently applied to atomic-spin ensembles to create  spin-ensemble states~\cite{Christensen2014} exhibiting significant non-classicality~\cite{McConnell2015}. These non-Gaussian operations can be used to create highly non-classical states and it has been theoretically shown that the addition operation creates non-classicality for any initial mean thermal occupation~\cite{Mandel1986, Agarwal1992, Kim2005}.

Cavity quantum optomechanics now provides a means to perform single-phonon addition and subtraction to macroscopic mechanical resonators~\cite{Vanner2013}. These operations were first demonstrated experimentally using optical phonon modes in bulk diamond and then using silicon photonic-crystal structures, where non-classical light-matter correlations~\cite{Lee2012, Riedinger2016}, and entangled states of two mechanical resonators~\cite{Lee2011, Riedinger2018} were generated. There is significant scope for further exploration of single-phonon addition and subtraction within optomechanics to, e.g. generate a wide range of macroscopic quantum states that are yet to be experimentally realized, and for new studies of quantum thermodynamics.

Curiously, when single-quantum addition or subtraction is applied to a thermal state, the mean number of quanta actually increases in both cases. Indeed, for a thermal state of mean occupation $\bar{n}$, when applying an addition (subtraction) operation, the mean occupation undergoes the transformation $\bar{n} \rightarrow 2\bar{n} + 1$ ($\bar{n} \rightarrow 2\bar{n}$). Though this increase appears counterintuitive, an understanding of this effect can be obtained by considering the Bayesian inference with the information gained by the measurement that heralds this non-unitary operation. This behaviour has been observed for thermal optical fields by performing heralded single-photon addition/subtraction followed by homodyne detection~\cite{Zavatta2007, Parigi2007}, and the approximate doubling of the mean occupation was utilized for work extraction in Ref.~\cite{Vidrighin2016}. Though these operations are now well studied for optical fields, they remain far less explored for other bosonic systems. In particular, the approximate doubling of the mean occupation by these operations to a thermal state is yet to be demonstrated for any system other than traveling light fields.

In this Letter, we report the observation of doubling of the mean occupation of a mechanical oscillator via heralded single-phonon addition and subtraction in a Brillouin-optomechanical system. We measure the temporal dynamics of the resulting increase in the mechanical excitation via a heterodyne detection scheme and observe the aforementioned doubling with a high signal-to-noise ratio. This work combines both photon counting and optical-dyne detection in a single experiment, thus taking a step towards the realization of hybrid quantum protocols that exploit both discrete and continuous variables~\cite{Milburn2016}. Moreover, the strong optomechanical coupling achievable in this geometry~\cite{Enzian2019} and the excellent mechanical coherence times achievable with crystalline materials~\cite{Galliou2013, Renninger2018} provides a promising experimental path for the development of quantum memories and repeaters.

\begin{figure*}[ht]
    \centering
    \includegraphics[width=0.9\linewidth]{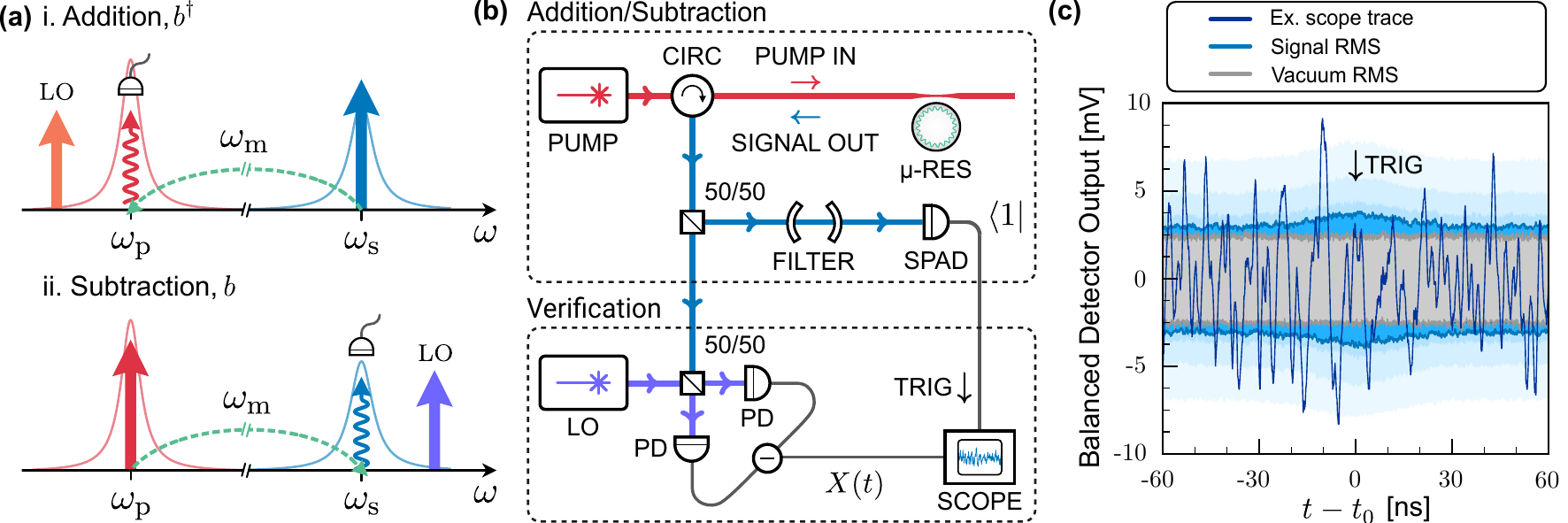}
    \caption{\small Scheme and setup. (a) Optical pumping and detection configuration comprising a pair of optical resonances spaced by approximately the mechanical frequency. For single-phonon addition, the higher-frequency mode of the pair is pumped and the down-shifted signal field is filtered and then detected by a single-photon counter to herald the operation. The frequencies of pumping and detection are reversed for the subtraction operation. (b) Schematic of the experimental setup that uses a pump laser and single-photon detection for the single-phonon addition/subtraction operations, and a separate laser for verification via heterodyne detection. (LO: local oscillator laser, PD: photodiode, SPAD: single photon avalanche photodiode.) (c) An example recorded heterodyne time trace about a subtraction event (dark blue curve). In the background, the optical vacuum, and signal RMS levels are plotted (grey, and blue, respectively), and a colour gradient is used to show the statistics of the signal feature above the optical vacuum noise.
    }
\end{figure*}

\textit{Brillouin single-phonon addition and subtraction scheme.}---Brillouin scattering is a nonlinear-optical process involving the interaction between two optical fields and one high-frequency acoustic wave. In this work, an optical pump field interacts with thermally excited phonons to create frequency down-shifted (Stokes) or up-shifted (anti-Stokes) fields of light. For the anti-Stokes case, the pump interacts with counter-propagating phonons and for the Stokes case, the pump interacts with co-propagating phonons. In both cases, these new optical fields are backscattered with respect to the pump field. In bulk materials, both the Stokes and anti-Stokes processes occur simultaneously and with similar strengths. To break this symmetry and select only one of these processes, we utilize two optical modes of a cavity approximately spaced by the mechanical frequency, see Fig.~1(a). By pumping the higher-frequency mode of the pair, anti-Stokes scattering can be strongly suppressed, allowing the Stokes scattering to be selectively driven. Similarly, by pumping the lower-frequency mode of the pair, the anti-Stokes process can be selectively driven. In this pumped regime, the interaction can be linearized and accurately described by quadratic interaction Hamiltonians. For the Stokes-scattering case, the interaction is modelled by a photon-phonon two-mode squeezing Hamiltonian $H/\hbar = G(a b + a^\dagger b^\dagger)$, while for the anti-Stokes case, the interaction is modelled by a photon-phonon beam-splitter Hamiltonian $H/\hbar = G(a^\dagger b + a b^\dagger)$. Here, $a$ is the annihilation operator for the optical cavity mode supporting the scattered field, $b$ is the mechanical annihilation operator, and $G$ is the interaction rate for these two separate cases, which is proportional to the intrinsic coupling rate and the intracavity pump amplitude.

In cavity optomechanics, performing single-phonon addition or subtraction by driving the blue or red sideband followed by single-photon counting of the scattered signal was first considered theoretically in Ref.~\cite{Vanner2013}, and we have adapted that approach for the Brillouin-scattering-based experiment we perform here. To implement a single-phonon addition operation, the higher-frequency optical cavity mode of the pair is weakly pumped by a coherent state (cf. Fig.~1(a)) to drive the two-mode-squeezing interaction and the resulting counter-propagating Stokes signal is separated from the pump and detected by a single-photon counter to herald the addition operation. For this operation, the angular frequency of the signal field is $\omega_\textrm{s} = \omega_\textrm{p} - \omega_\textrm{m}$, where $\omega_\textrm{p}$ refers to the pump laser angular frequency, and $\omega_\textrm{m}$ is the mechanical angular frequency. Implementing a single-phonon subtraction operation is performed in a similar manner where the lower-frequency mode of the pair is pumped to bring the beam-splitter interaction into resonance prior to single-photon detection. The signal field for this operation has the angular frequency $\omega_\textrm{s} = \omega_\textrm{p} + \omega_\textrm{m}$. In this weak-pump regime, the heating or cooling of the mechanical mode via the optomechanical interaction with the pump field is negligible. Moreover, the probability of two or more photons being detected is also negligible, ensuring that the measurement heralds a single-phonon operation. Under these conditions, we model these non-unitary operations to the input mechanical state with the mechanical creation and annihilation operators $b^\dag$ and $b$, respectively, and, in this weak-drive regime, these operations have a heralding probability proportional to the mechanical occupation $\bar{n}$. Brillouin optomechanics is very well suited to implementing single-phonon addition and subtraction owing to the high mechanical frequencies available in the back-scattering configuration, and the ability to be readily implemented in ultra-low optical loss materials. 

Also note that as we are pumping only in the forward direction and observing scattered light in the backward direction, due to the Brillouin phase-matching conditions, we are then applying the addition operation to a forward acoustic wave and the subtraction operation to a backward wave. Should it be desired to perform addition or subtraction to the same mechanical mode, the microresonator can be pumped in the reverse direction, as the system is symmetric under inversion of the propagation directions of all three waves.

Following an addition or subtraction operation, we use a heterodyne detection scheme to measure the variance of the mechanical amplitude fluctuations to verify the effect of the operation on the mechanical state. For experimental simplicity, we use the same continuous weak drive for the addition and subtraction operations and for the verification. For weak optical drive and $\bar{n}>1$, the optical amplitude on top of the vacuum noise on the Stokes or anti-Stokes scattered light serves as a proxy for the mechanical amplitude, thus enabling the dynamics of the mechanical fluctuations about the herald event to be characterised.


\textit{Experimental Setup.}---To experimentally implement this scheme, a BaF$_2$ optical microresonator fabricated using a diamond nanolathe is used. The device has a micro-rod-resonator geometry \cite{DelHaye2013} with a diameter of approximately $1.5$~mm, and a lateral confinement region with a radius of curvature of approximately $40$~$\mu$m. An optical microscope image of the microresonator is provided in the Supplementary Material~\cite{Supp}.

We employ an all fiber-based optical setup and use a continuous-wave pump laser running at $1550$~nm.  A silica tapered optical fiber is used to evanescently couple to the microresonator, and despite the small difference in refractive index between the fiber and the resonator, up to and beyond critical coupling to the optical cavity modes can be achieved. By recording the optical transmission spectra with a weak field, we observe optical whispering-gallery-mode resonances with an intrinsic quality factor of $Q \simeq 10^{8}$, and identify a pair of optical resonances that have a frequency separation approximately equal to the Brillouin shift of $\omega_\textrm{m} /2\pi = 8.21$~GHz. These optical modes are not spaced by the free-spectral range of the microresonator but have different transverse spatial profiles while still providing significant Brillouin optomechanical coupling. For the taper-coupling position used, the optical amplitude damping rates of the lower- and higher-frequency cavity modes are $6.8$~MHz, and $7.8$~MHz, respectively. The mechanical mode is a pseudo-longitudinal elastic whispering-gallery wave~\cite{Sturman2015} that fulfills the Brillouin phase matching conditions~\cite{Supp}. We perform the experiment at 300~K (stabilized), which corresponds to a mechanical mean occupation of $\bar{n} \simeq 760$. Based on the experimental observations described below, we estimate the mechanical amplitude decay to be $\gamma/2\pi = (17.0 \pm 3.2)$~MHz, which is expected to be largely from intrinsic material damping owing to the room-temperature operation.

We lock to either the higher- or lower-frequency optical mode by Pound-Drever-Hall laser-frequency stabilisation. The laser input-pump power is on the order of $1$~mW, leading to intracavity powers of $0.7$~W, and optomechanical coupling strengths $G/2\pi$ of typically $2$~MHz. We intentionally perform this particular experiment in the weak optomechanical coupling regime, where the coupling rate $G$ is much smaller than the optical amplitude decay rate $\kappa$ and the mechanical amplitude decay rate $\gamma$. For these parameters, the steady-state optomechanical cooling/heating to the mechanical oscillator is less than $1\%$~\cite{Supp}. Brillouin Stokes and anti-Stokes light backscattered from the cavity is coupled out through the taper, is separated from pump light via an optical circulator, and is then detected via a single-photon avalanche photodiode (SPAD) to herald the single-phonon addition and subtraction operations, see Fig.~1(b). Prior to single-photon detection, spurious residual pump photons are filtered using two fiber-based Fabry-Perot filters that have an intensity FWHM linewidth of $120$~MHz and a free-spectral range of $25$~GHz. The SPAD was operated with a quantum efficiency of $\sim12.5$~\%, a gate rate of $50$~kHz, a detector recovery time of $20~\mu\text{s}$ between gates to avoid afterpulsing, and a 8~ns effective gate duration. Note that the optical losses in our setup are large, however, this only affects the heralding probability and not the fidelity of the operation performed. This resilience to optical loss is because the dark-count rate 3~s$^{-1}$ is much lower than the signal rate 500~s$^{-1}$ and the probability of multi-photon detection is negligible, thus a detector click corresponds to high-fidelity addition or subtraction of a single phonon.

A detection event by the SPAD at time $t_0$ heralds the single-phonon operation and the output TTL signal triggers a digital storage oscilloscope to record a heterodyne time trace. An example heterodyne trace and the RMS noise is plotted in Fig.~1(c). We repeat this measurement approximately 20,000 times for both single-phonon addition and subtraction, which enables us to easily observe the increase in the mean occupation and the non-equilibrium dynamics around the herald event. We implement both single photon detection and heterodyne detection by splitting the signal field with a 50:50 beamsplitter, which simplifies the setup at the cost of a reduced heralding rate. To implement the heterodyne detection, a balanced detector and a local oscillator shifted by approximately 150~MHz with respect to the scattered light frequency, is used.

\begin{figure}[!h]
\includegraphics[width=0.9\linewidth]{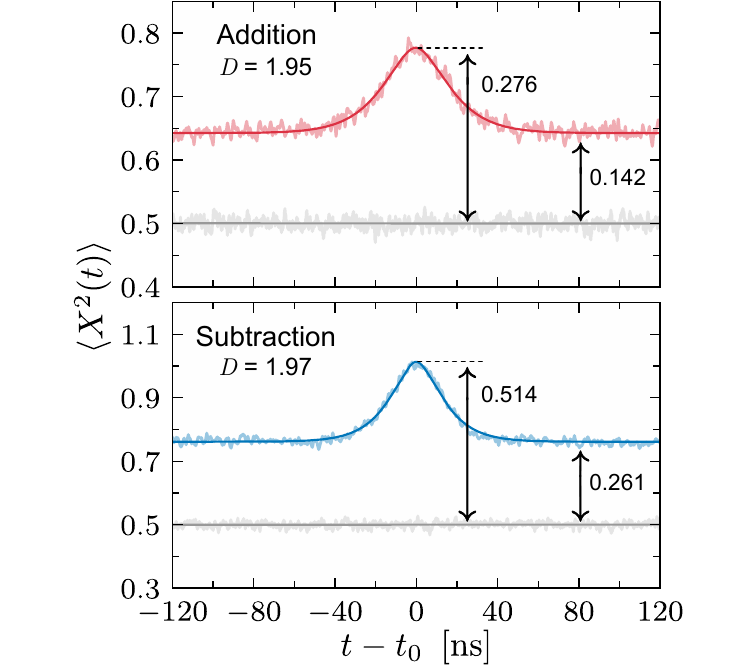} \\
\caption{\small Variance of the heterodyne signal as a function of time about the heralding event for the addition operation (red trace, above) and the subtraction operation (blue trace, below). The variance of the optical vacuum, measured separately, is normalised to $1/2$ and is shown in grey. The variance is determined from 20,000 runs of the experiment for each plot. About the heralding event $(t-t_0=0)$, a clear feature is observed indicating that the mechanical fluctuations have increased. Excellent agreement between the experimental data (pale lines) and theoretical fits (dark lines) is obtained~\cite{Supp}. The ratio of the variance above the vacuum noise at the center of the feature to far from the feature we label $D$, which is very close to 2 in both of these cases, indicating that the mechanical mean occupation has doubled to a high precision.
}
\label{Fig:Results}
\end{figure}

\textit{Results and Discussion.}---In Fig.~\ref{Fig:Results} the ensemble average of the square of the heterodyne signal $X$ is plotted with time about the herald event for both the addition and subtraction operations. This heterodyne signal is normalized such that $\langle X^2\rangle = 1/2$ when just optical vacuum is measured, which is shown as the grey lines in the two plots, and was experimentally verified to be 20~dB above the electronic noise of the detector. For both interactions used here, we experimentally verified that the scattered optical signal fields are rotationally invariant in phase space and hence did not phase-lock the local oscillator for the heterodyne verification measurements. For the single-phonon-addition operation, we observe a variance 0.142 above the vacuum level at a time far from the herald event, owing to the mechanical thermal fluctuations being mapped onto the optical field. Then, at the time of the herald event, we observe a feature in time that peaks at 0.276 above the vacuum level. These levels above the vacuum are proportional to the mechanical mean occupation and taking the ratio of the two indicates that the mean phonon number has increased by a factor of $D = 1.95 \pm 0.02$. For the single-phonon subtraction case,  we observed 0.261 above the vacuum away from the herald event and 0.514 above the vacuum level at the herald event, indicating that the mean occupation has increased by a factor of $D=1.97 \pm 0.02$. These factors are very close to the theoretical prediction of a doubling of the mechanical contribution to the quadrature variance.~\cite{Supp}. Note that the variances observed are different for the single-phonon-addition and subtraction cases as the pumping is swapped between the two optical resonances, which possess different linewidths, and the input pump power was also adjusted. Note also that the peak of the feature is reduced by approximately 1~\% due to small amount of darkcounts in the SPAD.

The temporal dynamics of the `doubling feature' are governed by the interplay between the optical and mechanical damping rates, $\kappa$ and $\gamma$, respectively. To quantitatively describe these dynamics we developed a model using Langevin equations for Brillouin optomechanical interactions, cavity input-output, and quantum measurement theory~\cite{Supp}. We find excellent agreement between the experimental data and the theoretical prediction with the only free fitting parameters in this model being the factor $D$ and the mechanical damping rate, with the latter being consistent with fits of the anti-Stokes heterodyne spectra. The estimates of these fitting parameters can be improved by taking into account the response of the optical filter prior to the SPAD and the spectral response of the balanced detector~\cite{Supp}.

A further contribution made by this work is that the measurement configuration introduced here can also be used to examine the degree to which multiple mechanical modes couple. This is possible as the presence of multiple mechanical sources will give rise to temporal interference fringes in the heterodyne signal around the herald event. A mathematical model for the time evolution of the quadrature variance in the presence of multiple mechanical modes is given in the supplementary material~\cite{Supp}. In this experimental work, as no temporal interference fringes are observed, we infer that our coupling is predominantly to a single mechanical mode.

\textit{Outlook.}---Brillouin optomechanics with high-frequency phonons~\cite{Enzian2019, Kharel2019} is emerging as a powerful new platform for quantum and classical optomechanics applications. Owing to the favourable properties of crystalline materials, such systems now provide excellent mechanical $Qf$ products~\cite{Galliou2013, Renninger2018}, where $Qf > 10^{17}$~Hz, hence enabling ultra-low mechanical decoherence even for modest cryogenic temperatures ($\sim$ 4~K). The BaF$_2$ crystalline resonators used here provide a promising path to access this regime for the whispering-gallery-mode geometry. Moreover, BaF$_2$ is an ultra-low-loss optical material well suited to these studies~\cite{Lin2014}, which enables optically-induced heating to be minimized and Brillouin optomechanical strong-coupling to be readily achieved~\cite{Enzian2019}. These advantages, combined with the techniques demonstrated here, open a rich avenue for further studies including quantum memories and repeaters~\cite{DLCZ2001, Sangouard2011}, which can be even combined with the backscatter operation of Brillouin scattering for (quantum) routing and non-reciprocity applications.

To the best of our knowledge this is the first optomechanics experiment that combines photon counting and optical dyne detection, which lays further groundwork for mechanical quantum-state engineering applications, such as mechanical superposition-state preparation via mechanical squeezing and then single-phonon addition/subtraction~\cite{Milburn2016}. Following a state-preparation protocol, one can then switch to a stronger anti-Stokes interaction and utilize the strong coupling achievable with these systems~\cite{Enzian2019} for mechanical state transfer to light for mechanical quantum-state reconstruction~\cite{Vanner2015}. Further exciting lines of study opened by this work, applicable to Brillouin-based and other optomechanical systems, include continuous-variable quantum-state orthogonalization~\cite{Vanner2013}, precision thermometry in Brillouin optomechanical systems when at low temperature, and utilizing the dramatic change in the mean occupation for measurement-based quantum thermodynamics applications with phonons.

\textit{Acknowledgements.}---We acknowledge useful discussions with J.~Clarke, T.~M.~Hird, I.~Galinskiy, M.~S.~Kim, K.~E.~Khosla, W.~S.~Kolthammer, G.~J.~Milburn, M.~Parniak, I.~Pikovski., E.~S.~Polzik, H.~Shen, and I.~A.~Walmsley. We also thank P. Del'Haye, J. Silver, A. Svela, and S. Zhang for assistance fabricating tapered optical fibers. This project was supported by the Engineering and Physical Sciences Research Council (EP/N014995/1, EP/P510257/1, EP/T001062/1), UK Research and Innovation (MR/S032924/1), the Royal Society, EU Horizon 2020 Program (847523 ‘INTERACTIONS’), and the Australian Research Council (CE170100012, FL150100019).

\clearpage

\onecolumngrid

\newcommand{\beq}{\begin{equation}}
\newcommand{\eeq}{\end{equation}}

\centering
    \large{\textbf{Single-Phonon Addition and Subtraction to a Mechanical Thermal State:}}
    \\ \vspace{2mm}
    \large{\textbf{Supplementary Material}}
    \normalsize{ }

\justify

\section{Single-phonon addition and subtraction to a thermal state}
%
    In this section we show how single-phonon addition and subtraction to a thermal state increases the mean occupation $\bar{n}$ according to:
    \begin{equation}
        \text{Addition:} \ \bar{n} \rightarrow 2 \bar{n} + 1 \qquad \text{Subtraction:} \ \bar{n} \rightarrow 2 \bar{n} \, .
    \end{equation}
%
\subsection{Single-phonon addition and subtraction}

In our experiment, we achieve single-phonon addition or subtraction by pumping the higher-frequency or the lower-frequency mode of a pair of cavity resonances, respectively, followed by single photon detection of the scattered signal that heralds the operation (see Figure 1 in the main text). For the case of addition, the pump field drives a two-mode-squeezing-type interaction, and the detected signal is red-shifted with respect to the pump. For the case of subtraction, the pump field drives a light-mechanics beam-splitter-type interaction, and the detected signal is blue-shifted with respect to the pump. These heralded processes are then close experimental approximations to the single-phonon creation and annihilation operators, $b^\dagger$ and $b$, respectively.

\subsection{Change in the mean thermal occupation}
    %
   Let's first compute the effect of single-phonon subtraction on an initial thermal mechanical state of motion. The density operator of the resulting mechanical state is obtained via
    \begin{equation}
        \rho_\text{sub} 
        =
        \frac{b \rho_\text{th} b^\dag }{\Tr(\rho_\text{th} b^\dag b)} 
        =
        \frac{b \rho_\text{th} b^\dag }{\bar{n}} \ ,
    \end{equation}
where $\rho_\text{th} = \sum_n  \frac{\bar{n}^n}{(\bar{n} + 1)^{n+1}}  |n\rangle \langle n |$ is the density operator of a thermal state. (For clarity, we have neglected here the proportionality factor that determines the heralding probability.) The mean phonon number of the single-phonon-subtracted state is then given by
    \begin{align}
        \langle b^\dag b \rangle_{-} 
        &= 
        \Tr(\rho_\text{sub} b^\dag b)  \\
        &= 
        \frac{1}{\bar{n}} \Tr(b \rho_\text{th} b^\dag b^\dag b) \\
        &= 
        \frac{1}{\bar{n}} \sum_n \frac{\bar{n}^n}{(\bar{n} + 1)^{n+1}} \langle n | b^\dag b^\dag b b | n \rangle \\
        &= 
        \frac{1}{\bar{n}} \sum_n \frac{\bar{n}^n}{(\bar{n} + 1)^{n+1}} n (n-1) \\
        &= 
        \frac{1}{\bar{n}(\bar{n} + 1)} \sum_n q^n n (n-1) \ ,
    \end{align}
where in the last line we introduced $q = \bar{n}/(\bar{n} + 1)$. We can now use $\frac{d^2}{dq^2} q^n = (n-1) n q^{n-2}$, and the sum of a geometric series, to write
    \begin{align}
        \langle b^\dag b \rangle_{-}
        &=    
        \frac{\bar{n}}{(\bar{n}+1)^3} \frac{d^2}{dq^2} \sum_n q^n \\
        &= 
        \frac{\bar{n}}{(\bar{n}+1)^3} \frac{d^2}{dq^2} \frac{1}{1-q} \\
        &=   
        \frac{\bar{n}}{(\bar{n}+1)^3}   \frac{2}{(1-q)^3} \\
        &=
        \frac{\bar{n}}{(\bar{n}+1)^3}   2 (\bar{n} + 1)^3 \\
        &= 
        2 \bar{n} \ .
    \end{align}

    The calculation for phonon addition, achieved using the two-mode-squeezing-type Hamiltonian, proceeds in a similar manner and results in $\langle b^\dag b \rangle_{+} = 2 \bar{n} + 1$.

\section{Dynamics}
In this section we derive the temporal evolution of the mean quadrature variance observed in the optical heterodyne detection around the single-phonon addition/subtraction event. For this purpose, we first solve quantum Langevin equations for the coupled optical and mechanical modes, and then compute the correlation between the single-phonon detection events that herald the addition/subtraction operations and the heterodyne signal.
\subsection{Brillouin Optomechanical Interaction}

The Brillouin optomechanical interaction in our system can be described by the three-mode Hamiltonian
\begin{equation}
    \frac{H_\text{int}}{\hbar} = g_0 (a_\text{blue} a_\text{red}^\dag  b^\dag +  a_\text{blue}^\dag  a_\text{red} b) \ ,
\end{equation}
where $g_0$ is the underlying optomechanical coupling rate, and $a_\text{red}$ and $a_\text{blue}$ are the optical mode operators associated with the two different optical modes of the cavity. The resonance frequencies of these two optical modes are approximately spaced by the frequency of an acoustic mode of the cavity, represented by mode operator $b$. The interaction rate $g_0$ depends on the  electrostrictive response of the medium, and the mode overlap between the two optical modes and the mechanical mode that participates.

For a sufficiently strong coherent drive applied to one of the optical resonances of the cavity, the quantum fluctuations of the drive field can be neglected, and we can approximate the field operator with a complex amplitude. For drive on the blue resonance one has $a_\text{blue} \rightarrow \alpha$ and the interaction Hamiltonian becomes
\beq
\frac{H_\text{TMS}}{\hbar} = G ( a_\text{red}^\dag  b^\dag +   a_\text{red} b)  \ , 
\eeq
where we have introduced the pump-enhanced optomechanical coupling rate $G = g_0 |\alpha|$. This Hamiltonian is a light-mechanics two-mode-squeezing (TMS) interaction, which has a form that has been well studied in the quantum optics community.

Analogously, when driving on the red resonance we find a light-mechanics beam-splitter Hamiltonian
\beq
\frac{H_\text{BS}}{\hbar} = G ( a_\text{blue}^\dag b + a_\text{blue} b^\dag) \ . 
\eeq
This Hamiltonian resembles that of an optical beam splitter, however, here the beam splitter is between light and sound, rather than between two optical fields. 

These two light-mechanics interactions utilized here are readily available in radiation-pressure-based optomechanics in the resolved-sideband regime by appropriate red or blue detuning. Thus, the protocol that we describe below can be employed in a very broad range of optomechanical systems.

\subsection{Quantum Langevin equations}
%
    The time evolution of the optical and mechanical modes, $a$ and $b$, respectively, including open-system dynamics, is modelled using quantum Langevin equations. 
    These equations read
    \begin{align}
        \label{eq:langevinequations}
        \dot{a} 
        &= 
        - i \, [a,H/\hbar] - \kappa \, a + \sqrt{2 \kappa} \, a_\text{in} \ , \\
       \label{eq:langevinequations_b}
        \dot{b} 
        &= 
        - i \, [b,H/\hbar] -\gamma \, b + \sqrt{2 \gamma} \, b_\text{in} \ ,
    \end{align}
    where $H$ is the optically driven light-mechanics Hamiltonian, $\kappa$ and $\gamma$ are the optical and mechanical amplitude decay rates, respectively, and $a_\text{in}$ and $b_\text{in}$ represent the optical and mechanical environments, respectively. These latter noise operators have the following correlations:
    \begin{align}\label{eq:noisecorrelation}
    \begin{split}
        \langle a_\text{in}^\dag (t) a_\text{in}(t') \rangle 
        &= 
        0 \ , \\
        \langle a_\text{in}(t) a_\text{in}^\dag(t') \rangle 
        &=
        \delta(t-t') \ ,\\
        \langle b_\text{in}^\dag(t) b_\text{in}(t') \rangle 
        &=
        \bar{n}_\text{th} \delta(t-t') \ ,\\
        \langle b_\text{in}(t) b_\text{in}^\dag(t') \rangle 
        &=
        (\bar{n}_\text{th} + 1) \delta(t-t') \ ,
    \end{split}
    \end{align}
where $\bar{n}_\text{th}$ is the mean occupation of the mechanical bath, the optical bath is at zero temperature, and the first moments as well as correlations between any other combinations of the two operators is equal to zero.

We will now solve the dynamics for single-phonon subtraction, which uses the beam-splitter Hamiltonian. The calculation for single-phonon addition, which uses the two-mode-squeezing interaction, proceeds in a similar manner. Working in a frame co-rotating with the optical and mechanical fields, we linearise the pump-field such that the Hamiltonian for the interaction takes the form of beam-splitter-type Hamiltonian $H = \hbar G (a b^\dag + a^\dag b)$. 
After inserting this Hamiltonian into Eqs. (\ref{eq:langevinequations}) and (\ref{eq:langevinequations_b}), we write the equations in matrix form in the frequency-domain:
    \begin{gather}\label{eq:solution1}
        \begin{pmatrix} 
        \widetilde{A}(\omega) \\ 
        \widetilde{B}(\omega) 
        \end{pmatrix} 
        = 
        \frac{1}{(i \omega + \gamma)(i \omega + \kappa) + G^2}
        \begin{pmatrix}  
        i \omega + \gamma &  -i G \\  -i G &  i \omega + \kappa 
        \end{pmatrix}   
        \begin{pmatrix}
        \sqrt{2 \kappa} \, \widetilde{A}_\text{in}(\omega)  \\
        \sqrt{2 \gamma} \, \widetilde{B}_\text{in}(\omega)
        \end{pmatrix} \ ,
    \end{gather}
where $\widetilde{A}$, $\widetilde{A}_\text{in}$,  $\widetilde{B}$ and $\widetilde{B}_\text{in}$, are the Fourier transforms of the operators $a$, $a_\text{in}$, $b$ and $b_\text{in}$, respectively.

Making the approximation that the coupling is weak, i.e. $G \ll \kappa,  \gamma$, the frequency-domain solution for the intra-cavity optical field simplifies to
    \begin{equation}
        \widetilde{A}(\omega) 
        =
        \frac{\sqrt{2 \kappa}}{i \omega + \kappa} \, \widetilde{A}_\text{in}(\omega) 
        - \frac{iG \sqrt{2 \gamma}}{(i \omega + \gamma)(i \omega + \kappa)} \, \widetilde{B}_\text{in}(\omega) \ .
    \end{equation}
Returning to the time-domain by applying the inverse Fourier transform, and using the convolution theorem, we arrive at
    \begin{align}\label{eq:BSsolution}
        a(t) 
        =
        \sqrt{2 \kappa} \,
        \Big( 
        e^{-\kappa t} \, \Theta(t) 
        \Big) 
        * a_\text{in}(t)
        -  
        \frac{i G \sqrt{2 \gamma}}{\kappa-\gamma} \,
        \bigg(
        \Big( e^{-\gamma t} - e^{-\kappa t} \Big) \, \Theta(t) \bigg) 
        * 
        b_\text{in}(t) \ ,
    \end{align}
    where $\Theta(t)$ is the Heaviside-Lorentz step function.

    Similarly, for the two-mode-squeezing-type Hamiltonian $H=\hbar G(a^\dag b^\dag + a b)$, one obtains
    \begin{align}\label{eq:TMSsolution}
        a(t) 
        =
        \sqrt{2 \kappa} \,
        \Big( e^{-\kappa t} \, \Theta(t) \Big) * a_\text{in}(t)
        -  
        \frac{i G \sqrt{2 \gamma}}{\kappa-\gamma} \,
        \bigg( \left( e^{-\gamma t} - e^{-\kappa t}\right) \, \Theta(t) \bigg) * b_\text{in}^\dag(t) \ .
    \end{align}
    Comparing Eqs.~(\ref{eq:BSsolution}) and (\ref{eq:TMSsolution}), we see that they have a very similar form, where the optical field operator $a$ is correlated to the mechanical noise operator $b_\text{in}$ for the beam-splitter interaction, and $a$ is correlated to $b_\text{in}^\dagger$ for the two-mode-squeezing interaction.

Note that for initial mechanical thermal states, which are invariant under rotations in phase space, we expect that the optical signal fields are also invariant in optical phase space according to our beam-splitter and two-mode-squeezer models, see Eq.~(\ref{eq:BSsolution}) and Eq.~(\ref{eq:TMSsolution}), respectively. We experimentally tested this phase independence by utilizing the narrow linewidth ($<~1$~kHz) of the pump and local oscillator lasers to perform phase-sensitive dyne detection during the $\sim 1$~ms phase coherence time and observed that the variance of the Stokes and anti-Stokes optical signals shows no dependence with phase with a high signal-to-noise ratio. Thus, we are able to perform the experiment without active phase stabilization of the local oscillator.

\subsection{Intra-cavity dynamics} \label{section:intra_cavity_dynamics}

\begin{figure}[htbp]
    \centering
    \includegraphics[width=80mm]{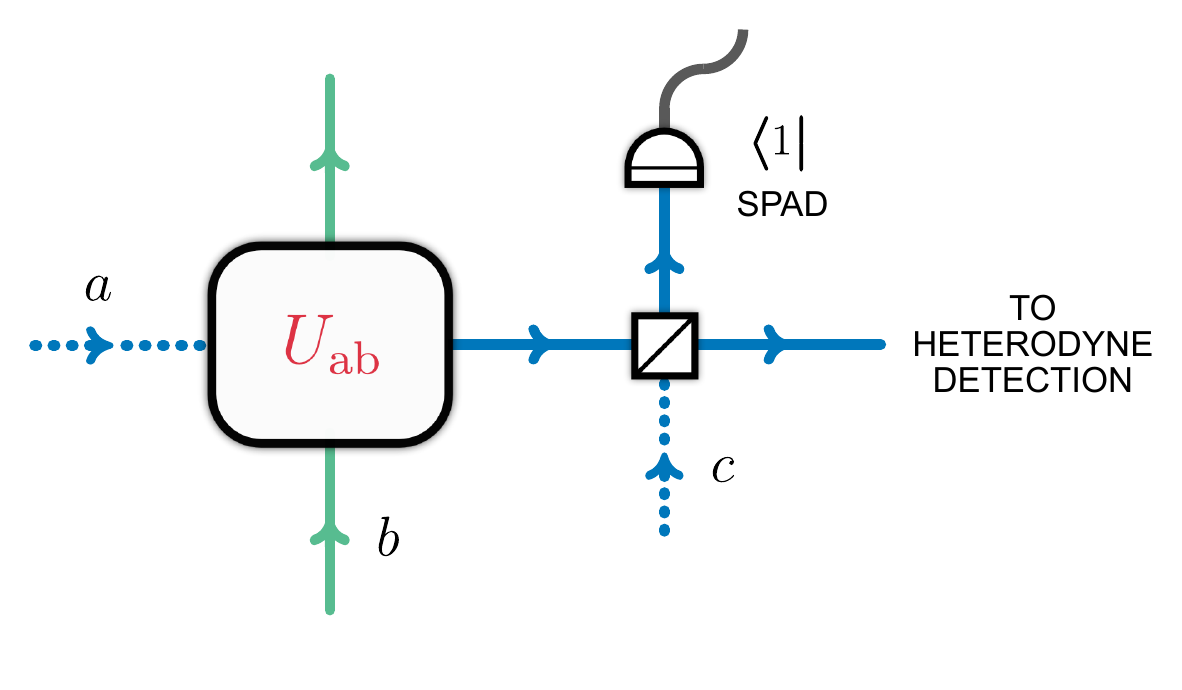}
    \caption{\small Simplified schematic of our experimental setup to implement and characterize single-phonon addition and subtraction to a mechanical thermal state. Here, $a$ represents the optical cavity mode of the scattered signal, $b$ is the mechanical mode, and $U_\text{ab}$ describes the (weak) light-matter beam splitter/two-mode-squeezing-type interaction. }\label{fig:toymodelmotivation}
\end{figure}

Using the solutions to the Langevin equations, we now compute the temporal evolution of the quadrature variance of the intra-cavity optical field around the time of single-phonon addition/subtraction events. This quantity provides a proxy for the mechanical quadrature variance, and is used to verify the effect of the operation on the mechanical oscillator. 

It is important to note at this stage that although it is insightful to consider the intra-cavity field, it is not experimentally accessible, and that cavity input-output relations, and an appropriate detector model, must be accounted for when considering the actual measurements. We omit this additional level of complexity for now, returning to it in Section \ref{sec:hetdet}.

In Fig.~\ref{fig:toymodelmotivation}, a simplified model for our joint click-dyne detection scheme is shown, where we have introduced an additional optical mode $c$, which participates in the optical beam-splitter between the single-photon counter and the heterodyne detector.
The quadrature variance of the optical cavity mode $X_\text{cav}^2$, for a general light-mechanics interaction, is then determined by computing
\begin{equation}
    \langle X_\textrm{cav}^2 \rangle 
    = \frac{1}{\mathcal{P}}  
    \Tr \left(
    {}_\text{c} \langle 1| \,
    B_\text{ac} U_\text{ab} \,\, 
    \rho_\text{m} \! 
    \otimes \!
    |0\rangle_\text{a} \langle 0| \!
    \otimes \!
    |0\rangle_\text{c} \langle 0| \,\,
    U_\text{ab}^\dag B_\text{ac}^\dag \,
    |1\rangle_\text{c} \,
    X_\text{cav}^2
    \right) \ ,
\end{equation}
where $U_\text{ab}$ represents the light-matter interaction, $B_{ac}$ the beam-splitter dividing the signal between the single-photon and heterodyne detectors, and $\mathcal{P}$ is the heralding probability. 

For a weak signal field $a$, and having negligible probability of two or more photons at the single-photon counter, we may write ${}_\text{c} \langle 1|B_\text{ac}|0\rangle_\text{c} = r \, a$, where $r$ is the reflectivity parameter of the beam splitter. This regime is relevant to our experiment and the cavity quadrature variance becomes
\begin{equation}
    \big\langle X_\textrm{cav}^2(\tau)  \big\rangle 
    =
    \frac{r^2}{\mathcal{P}} 
    \Tr \left( 
    \rho_\text{m} \!
    \otimes \!
    |0\rangle_\text{a} \langle 0| \,\,
    U_\text{ab}^\dag \,
    a^\dag(0)
    X_\textrm{cav}^2(\tau)
    a(0) \,
    U_\text{ab} 
    \right) \ .
\end{equation}

At this point we substitute the solutions to the Langevin equations, given by Eqs. (\ref{eq:BSsolution}) and (\ref{eq:TMSsolution}) respectively, to determine $X_\textrm{cav}^2(\tau)$, so as to take into account the full cavity dynamics. Note also that here we have used the time-stamp $t=0$ to indicate when the heralding event occurred and $\tau$ is the time after the heralding operation. The optical variance is then given by
\begin{equation}\label{eq:subtractedthermalvar}
    \big\langle X_\textrm{cav}^2(\tau) \big\rangle 
    =
    \frac{r^2}{\mathcal{P}}
    \big\langle a^\dag(0)  X_\textrm{cav}^2(\tau) a(0) \big\rangle 
    =
    \frac{
    \langle a^\dag(0)  X_\textrm{cav}^2(\tau) a(0) \rangle
    }{
    \langle a^\dag (0) a(0)\rangle
    } \ ,
\end{equation}
where we have used $\mathcal{P} = r^2 \langle a^\dagger(0) a(0) \rangle$.

Inserting $ X_\text{cav}(\tau) = \left( a^\dag(\tau) + a(\tau) \right) / \sqrt{2}$ into Eq. (\ref{eq:subtractedthermalvar}) we obtain 
\begin{equation}\label{eq:xmultipliedout}
        \big\langle  X_\textrm{cav}^2(\tau)  \big\rangle 
        =
        \frac{
        \langle a^\dag(0) a^\dag(\tau) a(\tau) a(0) \rangle   +    \langle a^\dag(0) a(\tau) a^\dag(\tau) a(0) \rangle
        }{
        2 \, \langle a^\dag (0) a(0) \rangle
        } \ .
\end{equation}
To explicitly see that the phase of the local oscillator does not influence the statistics of the measurement, one can replace $X_\text{cav}(\tau)$ in Eq. (\ref{eq:subtractedthermalvar}) with a general rotated optical quadrature $X_\theta(\tau) = (a(\tau) e^{-i\theta} + a^\dag(\tau) e^{i\theta}) / \sqrt{2}$ and note that the phase dependence drops out, i.e. 
\begin{equation}\label{eq:LOwayne}
        \big\langle  X_\theta^2(\tau)  \big\rangle 
        = \big\langle X_\textrm{cav}^2 (\tau)\big\rangle \ .
\end{equation}
Thus, we see that the heterodyne signal variance is unaffected by the local oscillator phase. (In our experiment, the local oscillator phase $\theta$ indeed advances at the constant rate $\theta = \omega_\text{het} t$ with respect to the signal field, where $\omega_\text{het}$ is the heterodyne frequency.)

To proceed, we apply the Isserlis-Wick theorem [L.~Isserlis, \textit{Biometrika} \textbf{12},
134–139 (1918) and S.~M.~Barnett, P.~M.~Radmore, \textit{Methods in theoretical quantum optics}, (1997)], which states that for Gaussian systems, the following is true:
\beq
\langle A B C D \rangle = \langle A B \rangle \langle C D \rangle + \langle A C \rangle \langle B D \rangle + \langle A D \rangle \langle B C \rangle \ , 
\eeq
where $A, B, C$ and $D$ correspond to (bosonic) operators.
Using this, we arrive at
\begin{align}\label{eq:thekey}
        \big\langle  X_\textrm{cav}^2(\tau)  \big\rangle     
        =    
        \frac{
        \langle a^\dag(0) a(\tau) \rangle 
        \langle a^\dag(\tau) a(0) \rangle
        }{
        \langle a^\dag (\tau) a(\tau) \rangle}
        +     
        \frac{
        \langle a^\dag(0) a(0) \rangle
        }{
        2 \langle a^\dag (\tau) a(\tau) \rangle
        }    
        \bigg(     
        \langle a(\tau) a^\dag(\tau) \rangle 
        + 
        \langle a^\dag(\tau) a(\tau) \rangle 
        \bigg)  \ , 
\end{align}
which, using the fact that the field leaving the cavity is stationary $\langle a^\dag(t_1) a(t_1) \rangle = \langle a^\dag(t_2) a(t_2) \rangle$, and that the two-time correlator has the symmetry $\langle a^\dag(0) a(t_1)\rangle = \langle a^\dag(t_1) a(0)\rangle^{*}$, we can further simplify to get 
\begin{equation}\label{eq:x2}
 \big\langle  X_\textrm{cav}^2(\tau) \big\rangle    
 =    
 \frac{
 |\langle a^\dag(0) a(\tau) \rangle|^2
 }{
 \langle a^\dag (0) a(0) \rangle}  
 +   
 \langle a^\dag(0) a(0) \rangle 
 + 
 \frac{1}{2} \ . 
\end{equation}

At this point, we have an expression that allows us to conveniently compute the quadrature variance for both single-phonon addition and subtraction operations (substituting in Eq.~(\ref{eq:BSsolution}) for the subtraction, and Eq.~(\ref{eq:TMSsolution}) for the addition). We also note that since $\langle a^\dag(0) a(0)\rangle$ is included in $\langle a^\dag(0) a(\tau)\rangle$, we only need to evaluate the latter.

Let's first compute the case of single-phonon subtraction.  As the optical vacuum noise has a mean photon number of zero (see Eq. (\ref{eq:noisecorrelation})), and cross terms between the optical and mechanical noise operators vanish, only the thermal mechanical noise terms contribute, and we obtain:
\begin{align}
\begin{split} \label{eq:a_0_a_tau_correlator}
\langle a^\dag(0) & a(\tau) \rangle = ... \\
... &= 
\frac{2 \gamma \, G^2}{(\kappa - \gamma)^2} 
\left\langle 
\int_{-\infty}^\infty \! dt' \! 
\int_{-\infty}^\infty \! dt'' \,
\bigg( 
\Big( e^{\gamma t'} - e^{\kappa t'} \Big) \, 
\Theta(-t')
\bigg) 
\bigg( 
\Big( e^{-\gamma(\tau - t'')} - e^{-\kappa(\tau - t'')} \Big) \, 
\Theta(\tau - t'')
\bigg) \,
b_\text{in}^\dag(t') \, b_\text{in}(t'') \right\rangle     \\
&= 
\frac{2 \gamma \, G^2}{(\kappa - \gamma)^2} 
\left\langle 
\int_{-\infty}^0  \!\!\! dt'  \!
\int_{-\infty}^\tau \!\!\!  dt'' \,
\Big( e^{\gamma t'} - e^{\kappa t'} \Big) 
\Big( e^{-\gamma(\tau- t'')} - e^{-\kappa(\tau-t'')} \Big) \, 
b_\text{in}^\dag(t') \, b_\text{in}(t'') \right\rangle \ .
\end{split}
\end{align}

We then make a case distinction, assuming first $\tau > 0$ :
\begin{align}\label{eq:a0atau_clunkyCorrelator}
\begin{split}
\langle a^\dag(0) a(\tau) \rangle_{\tau > 0} 
&=
\bar{n}_\text{th} \,
\frac{2 \gamma \, G^2 }{(\kappa-\gamma)^2} 
\int_{-\infty}^0 \!\!\!  dt' \,
\Big( e^{\gamma t'} - e^{\kappa t'} \Big) 
\Big( e^{-\gamma(\tau - t')} - e^{-\kappa(\tau-t')} \Big)  \\
&=   
\bar{n}_\text{th} \,
\frac{2 \gamma \, G^2 }{(\kappa-\gamma)^2} 
\bigg[ \, 
e^{-\gamma \tau} 
\int_{-\infty}^0 \!\!\!  dt'\,  
\Big( e^{2 \gamma t'} - e^{(\kappa+\gamma)t'} \Big) 
- 
e^{-\kappa \tau} 
\int_{-\infty}^0 \!\!\!  dt' \, 
\Big( e^{(\kappa+\gamma)t'} - e^{2 \kappa t'} \Big) \,
\bigg] \\
&= 
\bar{n}_\text{th} \,
\frac{2 \gamma \, G^2 }{(\kappa-\gamma)^2} 
\bigg[\, 
e^{-\gamma \tau} 
\bigg( 
\frac{1}{2\gamma} - \frac{1}{\kappa+\gamma} 
\bigg) 
- 
e^{-\kappa \tau} 
\bigg(
\frac{1}{\kappa+\gamma} - \frac{1}{2 \kappa}
\bigg) \,
\bigg] \\
&= 
\bar{n}_\text{th} \,
\frac{\gamma \, G^2 }{(\kappa + \gamma) (\kappa-\gamma)} 
\left( \frac{e^{-\gamma \tau}}{\gamma} - \frac{e^{-\kappa \tau}}{\kappa} \right) \ ,
\end{split}
\end{align}

and now $\tau < 0$, where for clarity, negative values of $\tau$ are written as $-|\tau|$:
\begin{align}
\begin{split}
\langle a^\dag(0) a(\tau) \rangle_{\tau < 0} 
&=  
\bar{n}_\text{th} \,
\frac{2 \gamma \, G^2 }{(\kappa-\gamma)^2} 
\int_{-\infty}^\tau \!\!\! dt'' \,
\Big(
e^{\gamma t''} - e^{\kappa t''} 
\Big) 
\Big(
e^{-\gamma \tau + \gamma t''} - e^{-\kappa \tau + \kappa t''} 
\Big)  \\
&= 
\bar{n}_\text{th} \,
\frac{2 \gamma \, G^2 }{(\kappa-\gamma)^2} 
\bigg[\, 
e^{\gamma |\tau|} 
\int_{-\infty}^{-|\tau|} \!\!\! dt'' \,
\Big( e^{2\gamma t''} - e^{(\kappa+\gamma)t''} \Big)
-
e^{\kappa|\tau|} 
\int_{-\infty}^{-|\tau|} \!\!\! dt'' \,
\Big( 
e^{(\kappa + \gamma)t''} - e^{2\kappa t''}
\Big) \,
\bigg] \\
&= 
\bar{n}_\text{th} \,
\frac{2 \gamma \, G^2 }{(\kappa-\gamma)^2} 
\bigg[\, 
e^{\gamma |\tau|} 
\bigg(
\frac{e^{-2 \gamma |\tau|}}{2 \gamma}  
- 
\frac{e^{-(\kappa+\gamma) |\tau|}}{\kappa+\gamma}  
\bigg) 
-  
e^{\kappa |\tau|} 
\bigg( 
\frac{e^{-(\kappa+\gamma)|\tau|}}{\kappa + \gamma}  
- 
\frac{e^{-2 \kappa |\tau|}}{2\kappa}  
\bigg) \,
\bigg] \\
&= 
\bar{n}_\text{th} \,
\frac{\gamma \, G^2 }{(\kappa+\gamma)(\kappa-\gamma)} \left( \frac{e^{-\gamma |\tau|}}{\gamma} - \frac{e^{-\kappa |\tau|}}{\kappa} \right) \ .
\end{split}
\end{align}

Thus, we can summarize the result for general $\tau$:
\begin{equation}\label{eq:generaladagaBS}
 \langle a^\dag(0) a(\tau)\rangle_- 
 =
 \bar{n}_\text{th} \,
 \frac{ \gamma \, G^2 }{(\kappa-\gamma)(\kappa+\gamma)} 
 \left( 
 \frac{e^{-\gamma |\tau|}}{\gamma} - \frac{e^{-\kappa |\tau|}}{\kappa} 
 \right) \ ,
\end{equation}
giving
\begin{equation}
\langle a^\dag(0) a(0)\rangle_- 
 = 
\bar{n}_\text{th} \,
\frac{G^2
}{
\kappa \, (\kappa + \gamma)}
 \ .
\label{eq:subtraction_intra_cavity_photon_number}
\end{equation}

For the case of single-phonon addition, the calculation proceeds in a similar manner using the result in Eq.~(\ref{eq:TMSsolution}), and we obtain
\begin{equation}
 \langle a^\dag(0) a(\tau)\rangle_+ 
 =  
 (\bar{n}_\text{th}+1) \,
 \frac{ \gamma \, G^2  }{(\kappa-\gamma)(\kappa+\gamma)} \left( \frac{e^{-\gamma |\tau|}}{\gamma} - \frac{e^{-\kappa |\tau|}}{\kappa} \right) \ ,
\end{equation}
and
\begin{equation}
\langle a^\dag(0) a(0)\rangle_+ 
 =  
 \big(
\bar{n}_\text{th} + 1 
\big)
\frac{G^2
}{
\kappa \, (\kappa + \gamma)} \ .
\end{equation}

Inserting these terms back into Eq. (\ref{eq:x2}), 
we obtain the expected quadrature variance in the optical cavity mode for a single-phonon-subtracted thermal state
\begin{equation}\label{eq:subX2}
    \big\langle  X_\textrm{cav}^2(\tau)  \big\rangle_{-} 
    = 
    \frac{1}{2} \  
    + \ 
    \frac{\bar{n}_\text{th} \, G^2}{\kappa(\kappa + \gamma)} \, 
    \Bigg( 
    1 \ 
    + \  
    \bigg(
    \frac{\kappa \, e^{-\gamma | \tau |}
    -
    \gamma \, e^{-\kappa | \tau |}
    }{
    \kappa - \gamma
    }
    \bigg)^2 \,
    \Bigg) \ ,
\end{equation}

and for a single-phonon added thermal state
\begin{equation}\label{eq:addX2}
    \big\langle  X_\textrm{cav}^2(\tau)  \big\rangle_{+} 
    = 
    \frac{1}{2} \  
    + \ 
    \frac{(\bar{n}_\text{th} + 1 ) \, G^2}{\kappa(\kappa + \gamma)} \ 
    \Bigg( 
    1 \ 
    + \ 
    \bigg(
    \frac{\kappa \, e^{-\gamma | \tau |}
    -
    \gamma \, e^{-\kappa | \tau |}
    }{
    \kappa - \gamma
    }
    \bigg)^2 \,
    \Bigg) \ .
\end{equation}

Examining Eq.~(\ref{eq:subX2}) we can identify three terms: the vacuum noise in the optical cavity mode, the equilibrium value of the transduced mechanical thermal noise, and a time-dependent non-equilibrium term corresponding to the effect of the single-phonon addition/subtraction operation. We see that at the time the operation is heralded ($\tau = 0$), the characteristic doubling of the mechanical quadrature variance is mapped onto the optical cavity variance.

Also note in Eqs~(\ref{eq:subX2}) and (\ref{eq:addX2}) that away from the herald event, i.e. $|\tau| \rightarrow \infty$, the background levels are proportional to $\bar{n}$ for the beam-splitter case, and $\bar{n}+1$ for the two-mode-squeezing case, and the expected sideband asymmetry, see e.g. [\textit{Phys. Rev. Lett.} \textbf{62}, 403 (1989); \textit{Rev. Mod. Phys.} \textbf{86}, 1391 (2014)], is obtained.

\subsection{Optomechanical cooling/heating}

Thus far we have assumed that the degree to which the mechanical mode is cooled (heated) by the application of the continuous-wave drive field on the red (blue) mode of the optical mode pair is negligible. Here we quantify the small amount of optomechanical cooling (heating) that occurs within the weak coupling regime as pertinent to this work.

Starting with the beam-splitter case, which corresponds to optomechanical cooling, and using the frequency-domain solution to the Langevin equations in Eq.~(\ref{eq:solution1}), the mean phonon number of the mechanical mode can be straightforwardly computed using
\begin{equation}
    \bar{n}_b = \frac{\bar{n}_\text{th}}{2\pi} \int_{-\infty}^{\infty}\! d \omega \,\, |\chi_{bb}(\omega)|^2 \ ,
\end{equation}
with the susceptibility $\chi_{bb}(\omega) = \frac{\sqrt{2 \gamma} (\kappa + i \omega)}{(\kappa + i\omega) (\gamma + i\omega) + G^2}$. Note that for the two-mode-squeezing case, which gives rise to optomechanical heating/amplification, one obtains $\chi_{bb}(\omega) = \frac{\sqrt{2 \gamma} (\kappa - i \omega)}{(\kappa - i\omega) (\gamma - i\omega) - G^2}$. 

Taking the limit of weak coupling, and expanding in $G^2$, we obtain expressions for the cooling and heating using the above. 
For cooling, the steady-state mean occupation is given by
\begin{equation}
    \bar{n}_\text{BS} = \bar{n}_\text{th} \left(1 - \frac{G^2}{\gamma (\kappa+\gamma)} + \mathcal{O}(G^4)\right) \ ,
\end{equation}
and for heating, the steady-state mean occupation is given by
\begin{equation}
    \bar{n}_\text{TMS} = \bar{n}_\text{th} \left(1 + \frac{G^2}{\gamma (\kappa+\gamma)} + \mathcal{O}(G^4)\right) \ .
\end{equation}
For our system parameters, which are summarised in Table (\ref{tab:sysparameters}) below, we obtain equilibrium phonon numbers of $\bar{n}_\text{BS} = 0.991~\bar{n}_\text{th}$, and $\bar{n}_\text{TMS} = 1.009~\bar{n}_\text{th}$, i.e. less than $1 \%$ of cooling/heating.

The steady-state occupations computed here can be thought of as a slightly shifted baseline starting occupation compared to the thermal equilibrium occupation, at the time when the addition or subtraction operation is applied.


\section{Detection}

\begin{figure*}[htp]
    \centering
    \includegraphics[width=150mm]{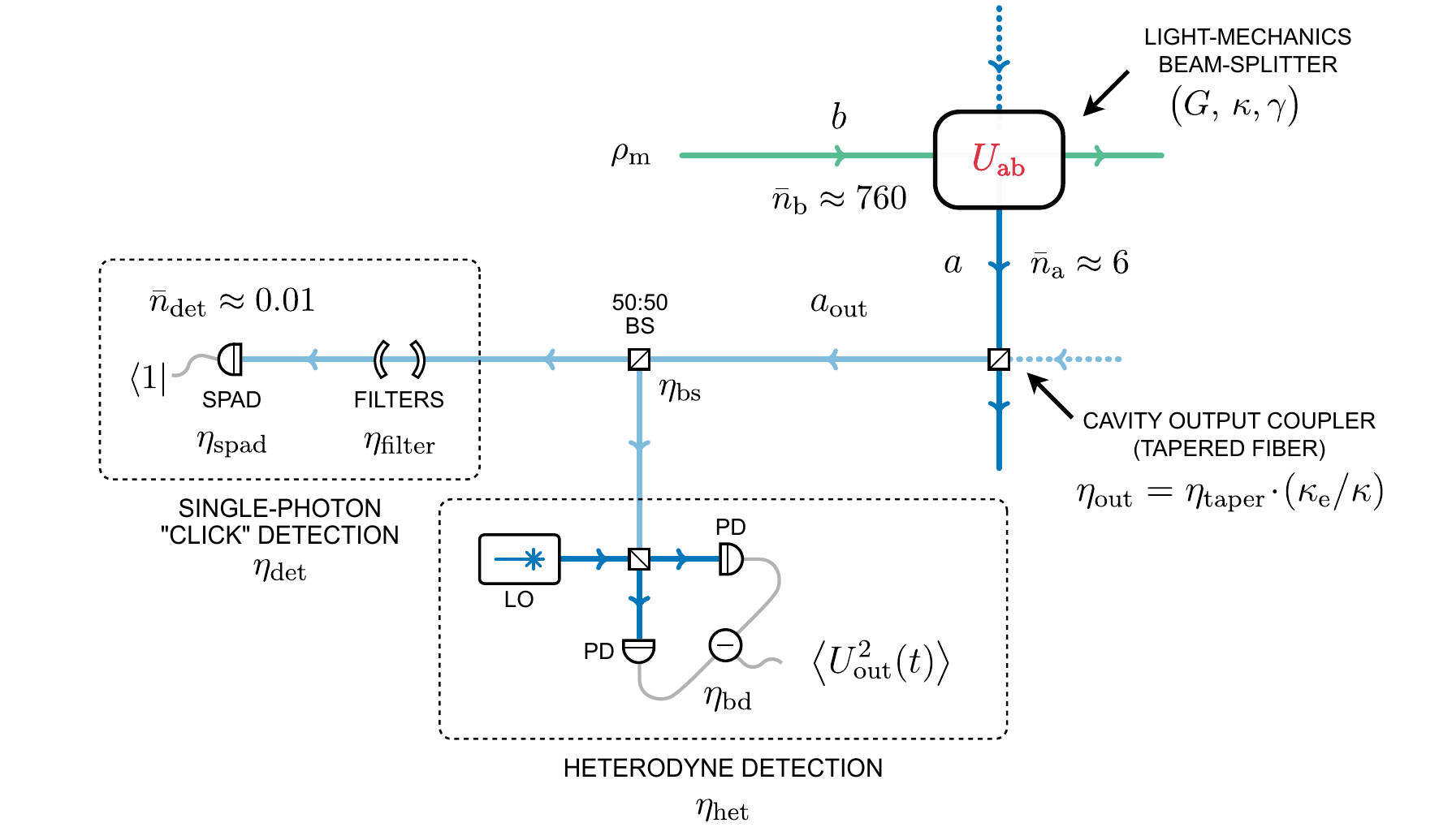}
    \caption{\small A model of the detection scheme used to herald, observe and verify single-phonon addition and subtraction to a mechanical thermal state. The mean number of equilibrium phonons in the mechanical mode $\bar{n}_\text{b}$, photons in intra-cavity optical mode $\bar{n}_\text{a}$, and photons per gate at the SPAD detector $\bar{n}_\text{det}$ are highlighted. Importantly, multi-photon detection events at the SPAD can be neglected, as the mean photon number per gate is much less than one.}
    \label{fig:toy_model_SPAD_detection}
\end{figure*}

As can be seen from Eqs. (\ref{eq:subX2}) and (\ref{eq:addX2}) when setting $\tau=0$, the doubling of the mechanical quadrature variance is mirrored in the intra-cavity field. We now discuss how input-output theory is applied to find the output field used to model the heralding operation via single-photon detection, and the heterodyne measurement. A model of this joint click-dyne detection scheme is depicted in Fig. \ref{fig:toy_model_SPAD_detection}.

\subsection{Heterodyne detection}\label{sec:hetdet}
%
We first demonstrate how heterodyne detection of the output signal field is intimately linked to the intra-cavity dynamics described in Section \ref{section:intra_cavity_dynamics}. Applying cavity input-output theory, the optical field emerging from the cavity is given by
 \begin{equation}\label{eq:outputmode}
        a_\text{out}(t)         =         a_\text{in}(t) - \sqrt{2 \kappa} \, a(t) \   ,
 \end{equation}
where it is important to note that $a_\text{in}$ and $a_\text{out}$ have units of $\text{s}^{-1/2}$, as opposed to the intra-cavity mode operator $a$, which is dimensionless. This field is then picked off by a 50:50 beam-splitter, before arriving at the heterodyne detection setup, as shown in Fig. \ref{fig:toy_model_SPAD_detection}.

\subsubsection{Output voltage signal of the balanced detector}

The output voltage signal of the balanced detector used for (finite-bandwidth) heterodyne detection in our experiment, can be modelled as
\begin{equation}
U_\text{out} (t)  
= 
\eta_\text{bd} \,  e  \, R_\text{fb} \,\,
\Big( 
\mathcal{F}^{-1} \,
\big\{
H(\omega)
\big\}
\Big)
*  
\frac{|\alpha_\text{LO}|}{\sqrt{2}}
\,
\Big( 
\hat{a}_\text{out}(t) \,  e^{-i \omega_\text{het} t} 
+ 
\hat{a}_\text{out}^\dag(t) \, e^{i \omega_\text{het} t} 
\Big) \ ,
\end{equation}
where $\eta_\text{bd}$ is the quantum efficiency of the photodiodes of the balanced photodetector, $e$ is the elementary charge, $R_\text{fb}$ and $H(\omega)$ are the feedback resistance and dimensionless transfer function of the transimpedance amplifier, respectively, $|\alpha_\text{LO}|$ is the coherent amplitude of the local osccilator, and $\omega_\text{het}/2\pi$ is the heterodyne frequency.
The low-pass behaviour of the balanced detector used in this experiment can be approximated by the transfer function $H(\omega) = 1/(1 + i \omega/\omega_\text{co})$, where $\omega_\text{co}$ is the characteristic cut-off frequency. 
Generalising for an arbitrary filter function, $h (t) = \mathcal{F}^{-1} \, \{ H(\omega) \}$, we can write the variance of the output-voltage signal of the balanced detector as
\begin{align} \label{eq:uoutsquared}
\big\langle 
U_\text{out}^2(t)
\big\rangle 
&= 
\frac{1}{2} \,
\big(
\eta_\text{bd} \, e \, R_\text{fb} \, |\alpha_\text{LO}|
\big)^2 
\,
\bigg\langle
\int_{-\infty}^{\infty}  \!\!\!  dt' \, 
h(t-t') \,
\Big( 
\hat{a}_\text{out}(t') + \hat{a}_\text{out}^\dag (t') 
\Big) 
\nonumber \\ 
&\qquad \quad \qquad \qquad \qquad \qquad \qquad
\times \ 
\int_{-\infty}^{\infty}  \!\!\!  dt'' \, 
h(t-t'') \,
\Big( 
\hat{a}_\text{out}(t'') + \hat{a}_\text{out}^\dag (t'') 
\Big)  
\bigg\rangle \ .
\end{align}

\begin{figure*}[htp]
    \centering
    \includegraphics[width=75mm]{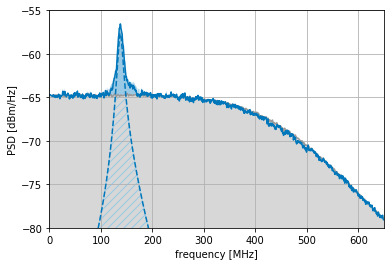}
    \caption{\small An example heterodyne spectrum showing the signal (blue) and vacuum noise floor (gray). The ``excess" vacuum noise is shaded in gray outside the blue hatched fill.}
    \label{fig:noises-illustration}
\end{figure*}

In our experiment, as the spectral width of the measured signal ($\sim15$ MHz) is much less than the bandwidth of the balanced detector ($\sim400$ MHz), and the heterodyne frequency ($\sim 150$ MHz) is much less than the cut-off frequency, i.e. the signal lies within a flat part of the detector's response, we are able to neglect any filtering effects that the finite-bandwidth response of the detector may have on the signal originating from the mechanics. Note, however, that the finite-bandwidth detector response cannot be neglected for the vacuum noise and is kept for these terms from here on. Furthermore, to ensure that the signal is sampled sufficiently, the output of the balanced detector is recorded with a digital storage oscilloscope with an analogue bandwidth of 1 GHz and a sample rate of 10~GSa/s. A plot of an example heterodyne spectrum illustrating the contributions of the thermal signal and the vacuum noise is depicted in Fig.~(\ref{fig:noises-illustration}).

\subsubsection{Vacuum noise}

The total vacuum noise contribution to the measured quadrature variance, is given by 
\begin{equation}
\begin{aligned}
\big\langle 
U_\text{out}^2(t) 
\big\rangle_\text{vac}
&= 
\frac{1}{2} \,
\big(
\eta_\text{bd} \, e \, R_\text{fb} \, |\alpha_\text{LO}|
\big)^2 \,
\int_{-\infty}^{\infty}  \!\!\!  dt' \, 
h^2(t-t') \\
&= 
\frac{1}{2} \,
\big(
\eta_\text{bd} \, e \, R_\text{fb} \, |\alpha_\text{LO}|
\big)^2 \,  
\int_0^{\infty}  \!\!\! d\omega \, |H(\omega)|^2 \ , 
\end{aligned}
\end{equation}
where we have used Plancherel's theorem to go from the first to the second line.

Note that as we are using a continuous-wave local oscillator, which does not match the spectral response of the cavity, `excess' vacuum noise is introduced into the heterodyne measurement, see Fig.~\ref{fig:noises-illustration}. (To improve the signal-to-vacuum-noise level in a future experiment, a pulsed pump and local oscillator could be employed that have a spectrum that matches the cavity response.) The ratio of signal to vacuum noise is therefore reduced at the detector. For our experiment, this meant that additional statistics were required in order to verify the effect of the single-phonon operations to the mechanical mode.

\subsubsection{Doubling factor}

We now derive an expression that contains the dynamics and characteristic doubling of the quadrature variance, that is similar to Eqs.  (\ref{eq:subX2}) and (\ref{eq:addX2}), for the optical output field.
Considering the case of subtraction, we substitute Eq. (\ref{eq:BSsolution}) into Eq. (\ref{eq:outputmode}), to obtain
\begin{equation}
\begin{aligned} \label{eq:outputfield}
a_\text{out}(t) \,
&= \,
a_\text{in}(t)
+
\frac{i G \sqrt{2\kappa} \sqrt{2 \gamma}}{\kappa-\gamma} \,
\bigg(
\Big( e^{-\gamma t} - e^{-\kappa t} \Big) \, \Theta(t) \bigg) 
* 
b_\text{in}(t)  \ ,
\end{aligned}
\end{equation}
for the field emerging from the cavity.
Using Eq.~(\ref{eq:outputfield}), and proceeding in a similar manner as Eqs. (\ref{eq:subtractedthermalvar}) to (\ref{eq:generaladagaBS}), we arrive at
\begin{equation}\label{eq:uoutsquaredExpl}
\begin{aligned}
\big\langle 
U_\text{out}^2(\tau) 
\big\rangle_- \, 
= \, 
&
\frac{1}{2} \,
\big(
\eta_\text{bd} \, e \, R_\text{fb} \, |\alpha_\text{LO}|
\big)^2 \,
\left[ \,
\int_0^{\infty}  \!\!\! d\omega \, |H(\omega)|^2 \ 
+
\eta_\text{out}
\frac{2 G^2}{\kappa + \gamma} \, 
\Bigg( 
\bar{n}_\text{th} 
+
\bar{n}_\text{th} \,
\bigg(
\frac{\kappa \, e^{-\gamma | \tau |}
-
\gamma \, e^{-\kappa | \tau |}
}{
\kappa - \gamma
}
\bigg)^2 \,
\Bigg) \right] \ ,
\end{aligned}
\end{equation}
for the variance of the output voltage signal of the balanced detector, where $\eta_\text{out} = \kappa_\text{e}/\kappa$ is a coupling efficiency. In practice this efficiency is multiplied with another efficiency $\eta_\text{taper}$, which accounts for additional losses, such as, for example, scattering out of the tapered fiber.
It is important to note here that we have neglected any filtering effect that the finite-bandwidth of the balanced detector has on the thermal contribution to the signal.

The equivalent expression for addition is found in a similar manner, and is given by

\begin{equation}\label{eq:uoutsquaredExpl_add}
\begin{aligned}
\big\langle 
U_\text{out}^2(\tau) 
\big\rangle_+ \, 
= \,
\frac{1}{2} \,
\big(
\eta_\text{bd} \, e \, R_\text{fb} \, |\alpha_\text{LO}|
\big)^2 \,
\Bigg[ \,
&
\int_0^{\infty}  \!\!\! d\omega \, |H(\omega)|^2
\, + \,
\eta_\text{out}
\frac{2 G^2}{\kappa + \gamma} \, 
\Bigg( 
\bar{n}_\text{th}
+  
(\bar{n}_\text{th} + 1) \,
\bigg(
\frac{\kappa \, e^{-\gamma | \tau |}
-
\gamma \, e^{-\kappa | \tau |}
}{
\kappa - \gamma
}
\bigg)^2 \,
\Bigg) \Bigg] \ .
\end{aligned}
\end{equation}

Once again, we can identify three terms in these expressions, corresponding to the optical vacuum noise, the steady-state mechanical noise contribution, and the dynamical term describing the action of the single-phonon addition/subtraction. 

It is important to note that, in contrast to Eqs.  (\ref{eq:subX2}) and (\ref{eq:addX2}), the optical vacuum noise in this case is generally not equal to $1/2$, as discussed in the previous section.
However, it is convenient to normalize Eqs. (\ref{eq:uoutsquaredExpl}) and (\ref{eq:uoutsquaredExpl_add}), such that the shot-noise level in the experiment does indeed correspond to $1/2$,  and we denote this normalized quantity as $X$, so as to indicate that the heterodyne measurement corresponds to the measurement of a field-quadrature. 
The normalized expression for single-phonon subtraction is given by

\begin{equation}\label{eq:subX2norm}
    \langle X^2(\tau) \rangle_{-} 
    = 
    \frac{1}{2} \ 
    + \ 
    \eta \, 
    \bar{n}_\text{th} \,\textbf{} 
    \Bigg( 
    1 \  
    +  \   
    \bigg(
    \frac{\kappa \, e^{-\gamma | \tau |}
    -
    \gamma \, e^{-\kappa | \tau |}
    }{
    \kappa - \gamma
    }
    \bigg)^2 \,
    \Bigg) \ ,
\end{equation}

and for single-phonon addition
\begin{equation}\label{eq:addX2norm}
    \langle X^2(\tau) \rangle_{+} 
    = 
    \frac{1}{2} \ 
    + \ 
    \eta \,
    ( \bar{n}_\text{th}+1) \,
    \Bigg( 
    1 \  
    +  \  
    \bigg(
    \frac{\kappa \, e^{-\gamma | \tau |}
    -
    \gamma \, e^{-\kappa | \tau |}
    }{
    \kappa - \gamma
    }
    \bigg)^2 \,
    \Bigg) \ ,
\end{equation}

where we have introduced the dimensionless parameter $\eta = \eta_\text{mm} \, \eta_\text{het}$, which takes into account both the total efficiency of the heterodyne detection $\eta_\text{het}$ (starting from the mechanical mode), and a mode mismatch parameter $\eta_\text{mm}$, which quantifies the `excess' shot noise due to the cw nature of the experiment. It is important to note that $\eta_\text{het}$ captures the efficiency of the effective light-mechanics beam-splitter ($\propto \! G^2$), and the tapered fiber output-coupler efficiency, as well as all other relevant detection efficiencies.


Using Eqs. (\ref{eq:subX2norm}) and (\ref{eq:addX2norm}), we can define a (doubling) factor $D_{\pm}$, which represents the increase in the mechanical contribution to the measured quadrature variance, relative to its thermal equilibrium value, upon heralded single-phonon addition or subtraction. For subtraction, we have
\begin{equation}
    D_{-} = \frac{\langle X^2(\tau = 0) \rangle_{-} - \frac{1}{2}}{\lim_{\tau \rightarrow \infty} \langle X^2(\tau)\rangle_{-} - \frac{1}{2}} = \frac{2 \bar{n}}{\bar{n}} = 2 \ ,
\end{equation}
and for addition,
\begin{equation}
    D_{+} = \frac{\langle X^2(\tau = 0) \rangle_{+} - \frac{1}{2}}{\lim_{\tau \rightarrow \infty} \langle X^2(\tau)\rangle_{+} - \frac{1}{2}} = \frac{2 (\bar{n} + 1)}{(\bar{n}+1)} = 2 \ .
\end{equation}
The calculation presented here was for the case of zero optomechanical detuning ($\Delta = 0$). It was checked that the characteristic doubling of the quadrature variance is unaffected by a finite optomechanical detuning, though these calculations are not presented here. The presence of a finite detuning, which is generally expected experimentally (if no further measures are taken), merely acts to reduce the effective interaction strength $G$. 

\subsection{Single photon detection events}

We now discuss the single-photon detection which heralds the addition/subtraction operation on the thermal mechanical state. It is important to note that the fidelity of the operation is only affected by the ratio of photo-counts-to-dark-counts, however, as the single-photon detectors employed in this experiment are unable to resolve photon number, it is essential that the detection mode has a low mean photon number. This ensures that the probability of multi-photon events is low, and thus detection of photon numbers greater than one can be neglected, making certain that we are indeed performing single-phonon addition and subtraction operations.

Using our detection model, as shown in Fig. \ref{fig:toy_model_SPAD_detection},  we will now demonstrate that a detection event at the SPAD can be confidently taken to correspond to a single-photon detection event, thus heralding a single-phonon operation.

For single-phonon subtraction, the rate of photons impinging on the detector for the continuous pump used here is computed with the aid of Eq.~(\ref{eq:subtraction_intra_cavity_photon_number}), along with the cavity input-output relations, and is given by
\begin{align}
\begin{split}
\label{eq:SPAD_rate}
    R_\text{det}  
    &= 
    \eta_\text{det} \, 
    \langle a_\text{out}^\dag(t) a_\text{out}(t) \rangle
    = 
    \eta_\text{det} \, 
    \bar{n}_\text{th} \,
    \frac{2  G^2}{\kappa+\gamma} \ ,
\end{split}
\end{align}
where $\eta_\text{det}$ is the total efficiency up until the point of single-photon detection in our experiment, which is explicitly 
\begin{align}
\begin{split}
\label{eq:SPAD_efficiency}
    \eta_\text{det}  
    &=
    \eta_\text{filter} \,
    \eta_\text{bs} \,
    \eta_\text{out} \,
\end{split}
\end{align}
where 
$\eta_\text{bs}$ is the efficiency of the beam splitter used for tapping off the light into two arms for single-photon and heterodyne detection,
$\eta_\text{filter}$ is the filtering efficiency,
and
$\eta_\text{out}$ is the overall taper coupling efficiency. 
The mean number of photo-counts at the detector (per gate) is then
\begin{align}
\begin{split}
\label{eq:SPAD_count}
    \bar{n}_\text{det}  
    &= 
    \eta_\text{spad} \, 
    R_\text{det} \, T_\text{gate}
    =
    \eta_\text{spad} \, 
    \eta_\text{det} \, 
    \bar{n}_\text{th} \,
    \frac{2  G^2}{\kappa+\gamma} \,
    T_\text{gate}
    \ ,
\end{split}
\end{align}
where $\eta_\text{spad}$ and $T_\text{gate}$ are the quantum efficiency and the effective gate length of the detector, respectively. 

For our experimental parameters (see Table \ref{tab:expparameters}), the overall detection efficiency is $\eta_\text{spad} \, \eta_\text{det} \approx 0.005$, and the number of photo-counts per gate is calculated to be of the order of $\bar{n}_\text{det} \approx 0.01$ for both addition and subtraction (noting that $\bar{n}_\text{th} \approx \bar{n}_\text{th} + 1$ for our system at room temperature). As  $\bar{n}_\text{det} \ll 1$, the assumption of single-photon events at the detector is safely valid.

\section{Effect of multiple mechanical modes}
\label{section:effect_of_multimode}
\begin{figure}[h!tbp]
\begin{center}
\includegraphics[width=60mm]{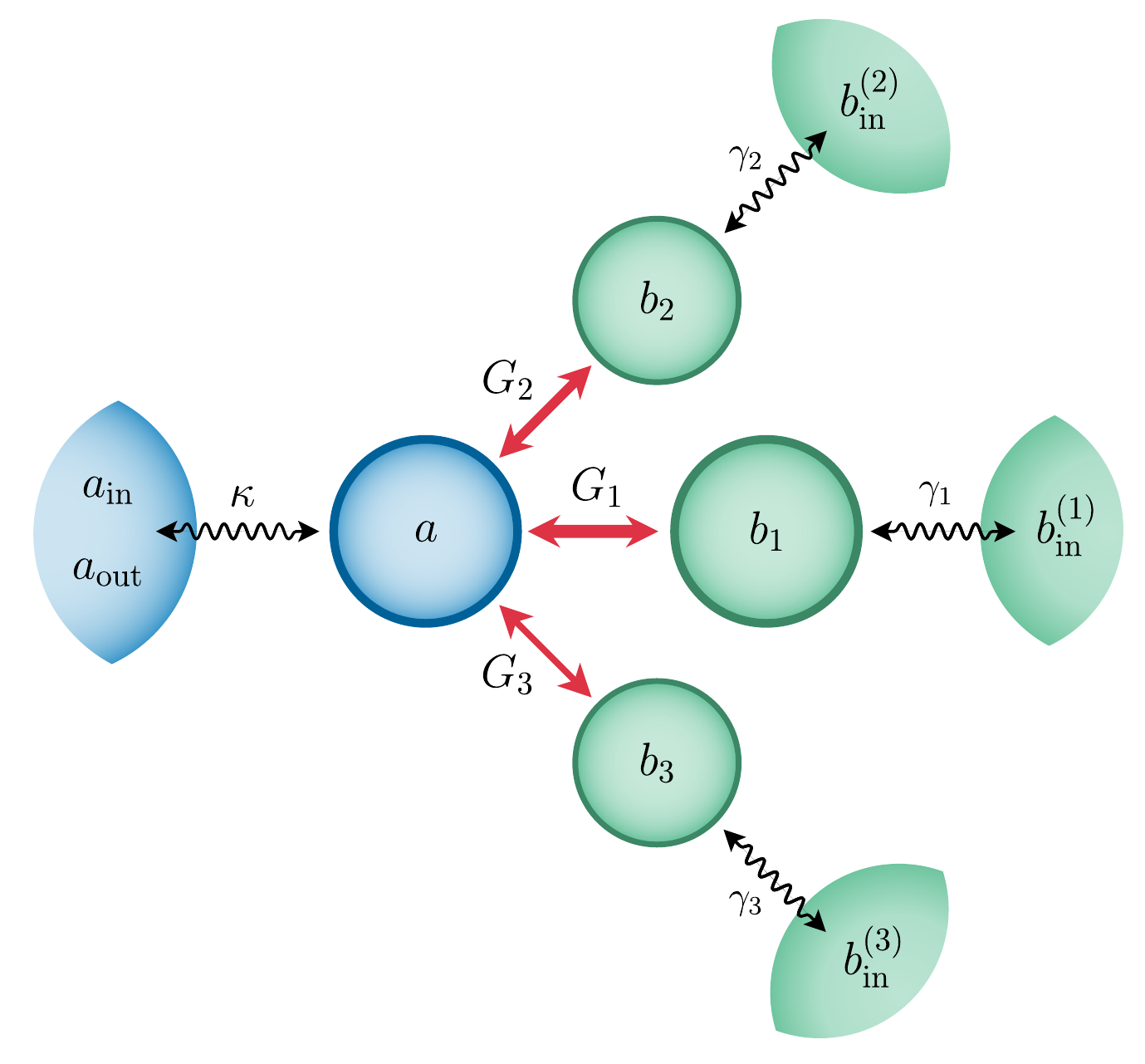}
\caption[Thermodynamic picture of multiple mechanical modes and reservoirs.]{Thermodynamic picture of multiple mechanical modes and reservoirs. Once a photocount is detected in the optical output mode, it can be thought of as having originated in one of multiple independent sources of mechanical thermal noise. The weighting of contributions is according to the scattering rate (mean photon number) in the output mode from each of the thermal reservoirs. Here, an example of three mechanical modes is depicted.}
\label{fig:thermodyn-multimode}
\end{center}
\end{figure}

In this section we discuss how the doubling signature of the quadrature variance under single-phonon addition/subtraction is changed in the scenario that the optomechanical interaction couples to multiple mechanical modes. This is shown schematically in Fig. \ref{fig:thermodyn-multimode}. In our whispering-gallery Brillouin-optomechanical system, the coupling rate is proportional to a triple overlap integral of the two optical and the mechanical mode field functions. Due to the axial symmetry of the system, the azimuthal integral corresponds to the condition that the azimuthal mode indices of the interacting waves are connected via $M_\text{m} = |M_\text{p}| + |M_\text{S}|$ for the coupling rate not to vanish, where $M$ is the azimuthal mode number, and m, p, and S stand for mechanical, pump and Stokes modes, respectively, and the absolute value was used to avoid definition ambiguity with respect to the propagation direction (we are considering counter-propagating optical waves in this work). The frequencies of the three involved fields, on the other hand, correspond to an individual value of the optomechanical detuning $\Delta = \omega_\text{p} - \omega_\text{S} - \omega_\text{m}$ for each combination of three modes. Achieving resonance in this Brillouin optomechanical system, i.e. a small detuning $\Delta$, is relaxed by the finite damping in both the optical and mechanical modes.

Now, for a given pair of optical resonances, a finite optomechanical coupling may be present for more than one mechanical mode that all share the same azimuthal mode index, but differ in their transverse (polar, radial) structure. Note that both the transverse overlap integral with the optics and the individual detunings due to waveguide dispersion lead to different effective coupling rates to these modes. In our system, a situation with predominantly single-mode coupling can be achieved by careful selection of the optical mode pair.

The optical and mechanical resonance structure can also be understood as a filter on an approximately spectrally flat white-noise heat bath, and thus it is clear that a non-Lorentzian optical density of states, e.g. through the presence of more than one optical cavity mode (cf. Fig. \ref{fig:thermodyn-multimode}), would result in a similar argument to what follows. However, the optical mode structure of the cavity can be directly accessed by transmission measurements with a scanning laser source, and it was in this way confirmed that the optical density of states was well described by a single optical cavity resonance for both the pump and the scattered modes.

In order to determine the effect on the doubling signature, we now consider two mechanical modes with damping rates $\gamma_1$ and $\gamma_2$, and detunings $\Delta_1$ and $\Delta_2$. As we shall see, while the doubling feature persists also in the multi-mode case, the temporal evolution of the mean quadrature variance shows an interference effect due to the spectral composition of the field.


\subsubsection{Heterodyne signal variance}

From a two mechanical mode beam-splitter type Hamiltonian in a rotating frame
\beq
\frac{H}{\hbar} = G_1 (a b_1^\dag + a^\dag b_1) + G_2 (a b_2^\dag + a^\dag b_2) - \Delta_1 b_1^\dag b_1 - \Delta_2 b_2^\dag b_2
\eeq
we can derive the quantum Langevin equations of motion for the optical and mechanical degrees of freedom
\begin{align}
\begin{split}
\dot{a} &= - i G_1 b_1 - i G_2 b_2  - \kappa a + \sqrt{2 \kappa} a_\text{in} \\
\dot{b}_1 &= -i G_1 a - (\gamma_1 + i \Delta_1) b_1 + \sqrt{2\gamma_1} b_\text{in}^{(1)} \\
\dot{b}_2 &= -iG_2 a - (\gamma_2 + i \Delta_2) b_2 + \sqrt{2\gamma_2} b_\text{in}^{(2)} \ .
\end{split}
\end{align}

After the application of the Fourier transform we obtain the equations in the frequency domain for the Fourier transforms of the mode operators
\begin{equation}
\begin{pmatrix} i\omega + \kappa & i G_1 & i G_2 \\ i G_1 & i\omega + \gamma_1 + i \Delta_1 & 0 \\ i G_2 & 0 & i \omega + \gamma_2 + i\Delta_2 \end{pmatrix} \begin{pmatrix} \tilde{A}(\omega) \\ \tilde{B}_1(\omega) \\ \tilde{B}_2(\omega) \end{pmatrix} = \begin{pmatrix} \sqrt{2\kappa} \tilde{A}_\text{in}(\omega) \\ \sqrt{2 \gamma_1} \tilde{B}_\text{in}^{(1)}(\omega) \\ \sqrt{2 \gamma_2} \tilde{B}_\text{in}^{(2)}(\omega) \end{pmatrix} \ .
\end{equation}

Inverting the matrix leads us to a general expression for the mode operator's spectral densities in terms of the input noise operators. After some algebra, we obtain for the optical mode operator's spectral density (focusing again on the mechanical noise contribution)
\begin{align}
\begin{split}
\tilde{A}(\omega) = ... \times \tilde{A}_\text{in}(\omega) \ 
&- \ \frac{i\sqrt{2 \gamma_1} G_1 (\omega - i \Gamma_2)}{G_1^2(\omega-i\Gamma_2) + G_2^2(\omega-i\Gamma_1) - (\omega-i\kappa)(\omega-i\Gamma_1)(\omega-i\Gamma_2)} \tilde{B}_\text{in}^{(1)}(\omega) \\
&- \ \frac{i\sqrt{2 \gamma_2} G_2 (\omega - i \Gamma_1)}{G_1^2(\omega-i\Gamma_2) + G_2^2(\omega-i\Gamma_1) - (\omega-i\kappa)(\omega-i\Gamma_1)(\omega-i\Gamma_2)} \tilde{B}_\text{in}^{(2)}(\omega) \ ,
\end{split}
\end{align}
where we introduced the shorthand $\Gamma_i = \gamma_i + i \Delta_i$.

In order to obtain the time domain solution we perform the inverse Fourier transform. The integration is non-trivial for the general case, but becomes straightforward if we make the weak coupling approximation (as we did previously in the single-mode scenario), i.e. $G_1, G_2 \ll \kappa, \gamma_1, \gamma_2$, in which case we neglect the corresponding terms in the denominator. The time domain solution for the optical mode operator then reads
\begin{align}\label{eq:atwomode}
\begin{split}
a(t) = ... * a_\text{in}(t) \ &+ \  i G_1 \sqrt{2\gamma_1} \frac{1}{\kappa-\gamma_1-i \Delta_1} \left(\left(e^{-(\gamma_1 + i \Delta_1)t} - e^{-\kappa t}\right) \Theta(t) \right) * b_\text{in}^{(1)}(t) \\
&+  i G_2 \sqrt{2\gamma_2} \frac{1}{\kappa-\gamma_2-i \Delta_2} \left(\left(e^{-(\gamma_2 + i \Delta_2)t} - e^{-\kappa t}\right) \Theta(t) \right) * b_\text{in}^{(2)}(t) \ .
\end{split}
\end{align}
As can be seen from the above, the optical and mechanical cavities act like filters on the (assumed) spectrally flat thermal bath sources. It is assumed that the presence of multiple optical frequency eigenmodes should have a similar filtering effect.

We can then write down the correlator (assuming that the Isserlis-Wick theorem still applies) associated with single phonon subtraction:
\beq\label{eq:x2multimode}
\langle X^2(\tau)\rangle = \frac{| \langle a^\dag(0) a(\tau) \rangle|^2}{\langle a^\dag(0) a(0) \rangle} + \langle a^\dag(0) a(0)\rangle + \frac{1}{2} \ .
\eeq

In the weak coupling approximation we can write  Eq. (\ref{eq:atwomode}) in the reduced form
\beq
a(t) = ... * a_\text{in}(t) + F_1(t;\kappa,\gamma_1,\Delta_1,G_1)*b_\text{in}^{(1)}(t) + F_2(t;\kappa,\gamma_2,\Delta_2,G_2)*b_\text{in}^{(2)}(t) \ ,
\eeq
where 
\begin{equation}
  F_i(t; \kappa, \gamma_i, \Delta_i, G_i) =  \frac{i G_i \sqrt{2\gamma_i}}{\kappa-\gamma_i-i \Delta_i} \left(\left(e^{-(\gamma_i + i \Delta_i)t} - e^{-\kappa t}\right) \Theta(t) \right)  
  \ .
\end{equation}

Accordingly, in order to determine the time evolution of the quadrature variance upon single-phonon subtraction, we must only compute the following
\begin{align}
\begin{split}
\langle a^\dag(0) a(\tau) \rangle &= \bar{n}_\text{th} \int dt' \left( F_1^{*}(-t') F_1(\tau-t') + F_2^{*}(-t') F_2(\tau-t') \right) \ .
\end{split}
\end{align}

\begin{figure}[htp]
\begin{center}
\includegraphics[width=80mm]{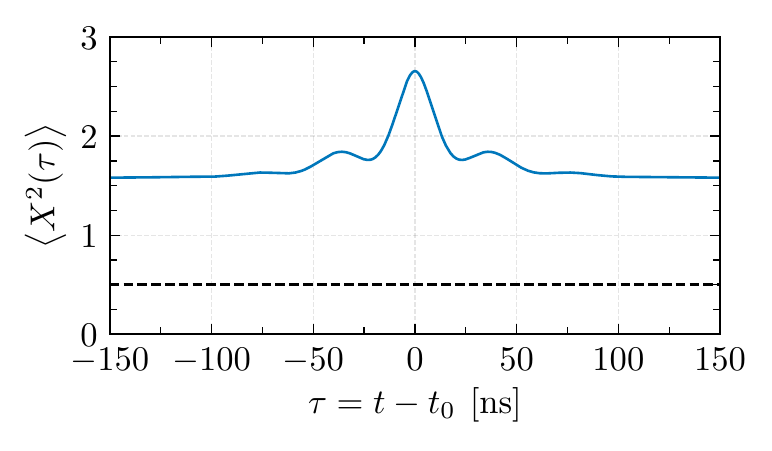}
\caption[Correlation function for a multi-mode field.]{Heterodyne signal variance in the multi-mode case for single-phonon subtraction (``click-dyne correlator''). The dashed black line represents vacuum noise. The parameters are given by the mechanical linewidths of $2\gamma_1/(2\pi) = 6~\text{MHz}$, $2\gamma_2/(2\pi) = 8~\text{MHz}$, an optical linewidth of $2 \kappa/(2\pi)$ = 50~MHz and detunings of $\Delta_1/(2\pi)$ = -15~MHz and $\Delta_2/(2\pi)$ = +10~MHz. The coupling rates are $G_1/(2\pi) = 1~\text{MHz}$ and $G_2/(2\pi) = 0.5~\text{MHz}$. }
\label{fig:multimodeg2plot}
\end{center}
\end{figure}

Figure \ref{fig:multimodeg2plot} shows a plot of the quadrature variance in a multi-mode scenario according to Eq. (\ref{eq:x2multimode}), where we considered two mechanical modes with linewidths of 6 and 8 MHz, respectively, an optical linewidth of 50 MHz and detunings of -15 and +10 MHz. The relative rates $\propto G_i^2$ have a ratio of 4 to 1. 
%

\section{Experimental System Parameters} \label{section:experiment}

\begin{figure*}[htp!]
    \centering
    \includegraphics[width=120mm]{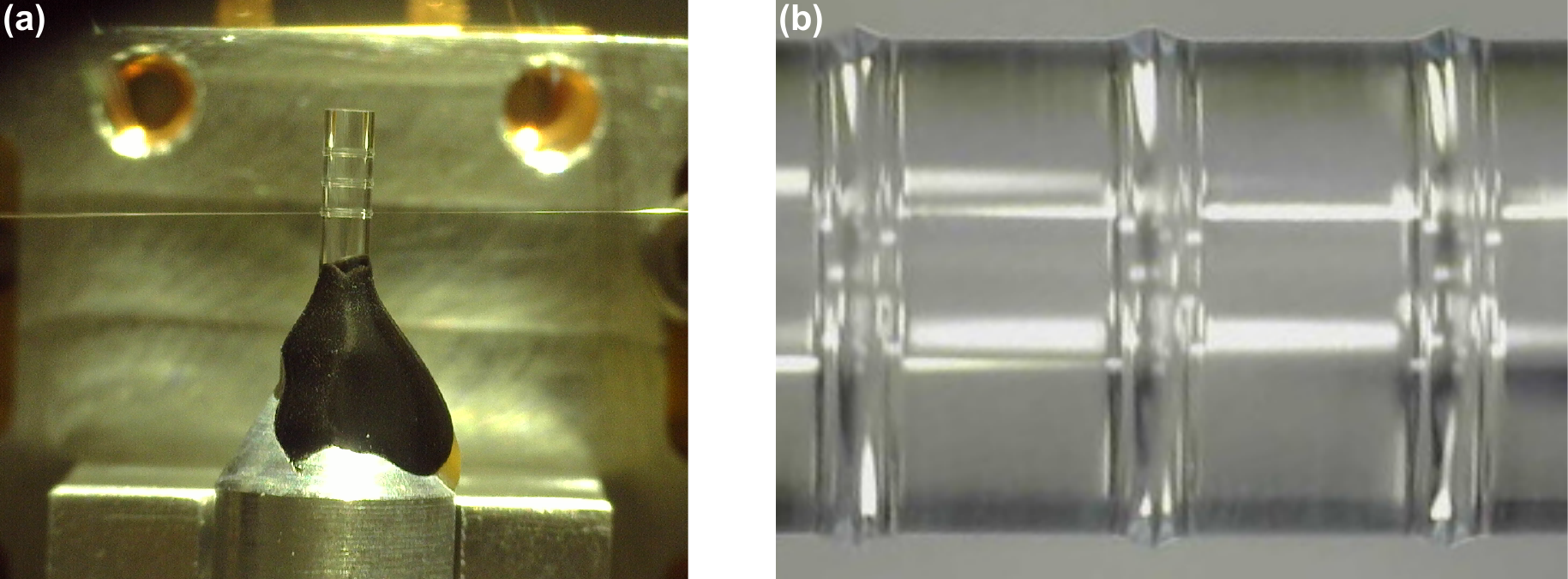}
     \caption{\small Optical microscope images of the $\text{BaF}_2$ microrod resonator as used in the experiments. (a) Resonator and accompanying fused-silica tapered optical fiber. (b) Close-up of the resonator. The rod has three regions cut into it which support whispering-gallery modes. These `bulges' have a diameter of approximately $1.5$ mm, and a lateral confinement region with a radius of curvature of approximately $40~\mu\text{m}$. }
     \label{fig:baf2_resonator}
\end{figure*}

\begin{figure*}
\centering
\includegraphics[width=120mm]{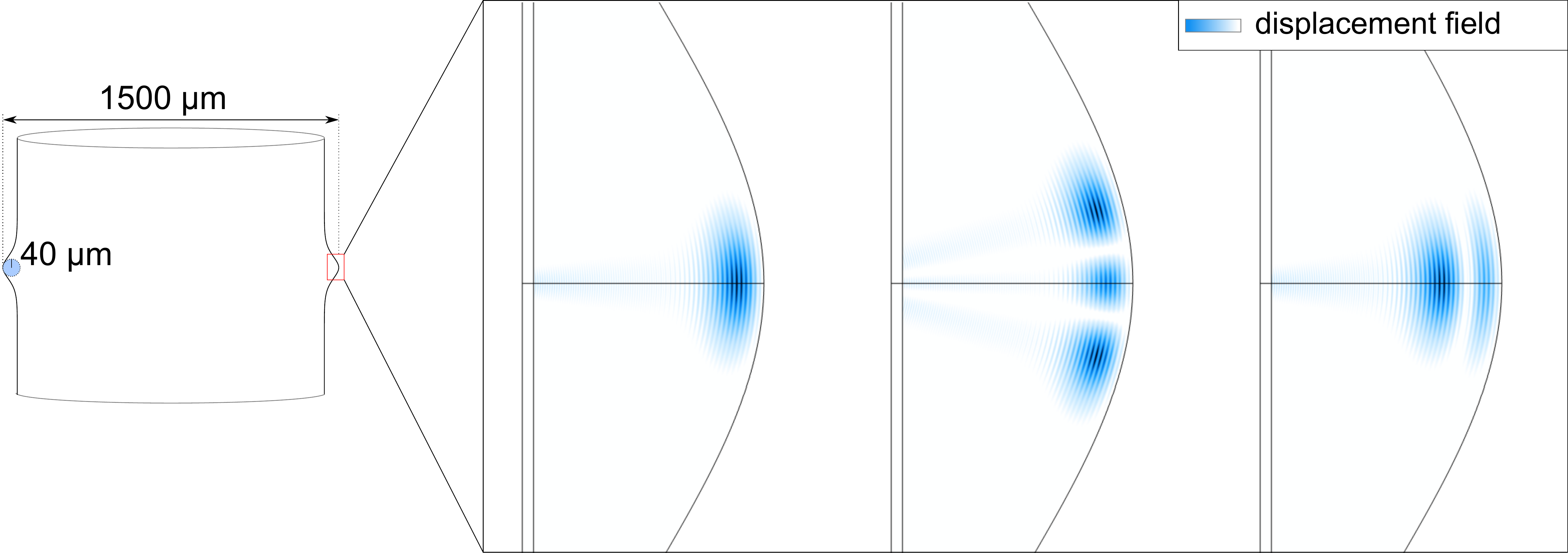}
\caption{\small Left part: Diagram of the micro-rod-resonator and its (approximate) dimensions. Right part: Simulated mechanical-mode profiles for the fundamental pseudo-longitudinal mode (8.202~GHz, left), a higher-order mode with polar structure (8.213~GHz, middle), and a higher-order mode with radial structure (8.228~GHz, right).}\label{fig:mode_profiles}
\end{figure*}

Summaries of our system and experimental parameters are given in Table~\ref{tab:sysparameters} and Table~\ref{tab:expparameters}, respectively, and an optical microscope image of the BaF2 micro-rod-resonator is given in Fig.~\ref{fig:baf2_resonator}. To help further understand the type of mechanical mode coupled to in this experiment, in Fig.~\ref{fig:mode_profiles} we plot finite-element simulations of the mechanical mode profiles of these elastic-wave whispering-gallery modes for the micro-rod-resonator geometry used here. For the resonator diameter (approximately 1.5 mm), the azimuthal mode number of the mode addressed in the experiment is $M \approx 8900$. While the transverse structure of this mode cannot be currently identified in this experiment with confidence, our simulations return eigenfrequencies for low transverse mode numbers (depicted) that are consistent with the experimentally observed value of 8.21~GHz.

An outstanding and important technical challenge in the optomechanics community at present is the development of a platform which simultaneously provides a high mechanical $Qf$ product, has low optical losses and absorption, and enables highly selective beam-splitter or two-mode-squeezer coupling to a single mechanical mode. The system used in this work achieves the latter two points, but low mechanical decay rates are not possible at room temperature owing to the material properties. We have identified three contributing factors to the total mechanical decay rate: surface roughness, material damping, and geometry-dependent radiative losses. As discussed in the main text, intrinsic material damping is expected to significantly reduce in crystalline barium fluoride at cyrogenic temperatures providing significant promise to simultaneously achieving these three points. Furthermore, owing to the similarities between the mechanical and optical whispering-gallery-modes utilized in this system, we expect that the contributions to the mechanical damping due to surface roughness and geometric damping can be made very small as very high optical $Q$ factors are achieved ($Q>10^8$).

While in the current room-temperature experiment, identification of the mechanical mode was difficult, with narrower mechanical linewidths at cryogenic temperature it should be easier to match the observed mechanical resonances with the eigenfrequencies obtained via finite-element simulation. This will enable identification of the mechanical mode numbers and transverse mode profile.


\begin{table}[htp!]
    \centering
        \caption{System parameters.}
    \label{tab:sysparameters}
    \begin{tabular}{|c|c|}
         \hline
         \textbf{Parameter} & \textbf{Value} \\
         \hline
         Resonator diameter, $D_\text{res}$ & 1.5 mm \\
         Pump wavelength, $\lambda_\text{p}$ & 1550 nm \\
         Mechanical frequency, $\omega_\text{m}/2\pi$ & 8.21 GHz \\
         \hspace{5pt} Optical linewidth (FWHM), $2\kappa_1/2\pi, 2\kappa_2/2\pi$ \hspace{5pt} & 13.5~,~15.5 MHz \\ 
         Mechanical linewidth, $2\gamma/2\pi$ & $(34.0  \pm 6.4 ) \ \text{MHz}$ \\
         Optomechanical coupling rate, $G/2\pi$ & $\sim$ 2 MHz \\  
         Mean phonon number, $\bar{n} \rightarrow 2 \bar{n}$ & \hspace{5pt} $760 \rightarrow 1520$ \\
         \hline
    \end{tabular}
\end{table} \,

\begin{table}[htp!]
    \centering
        \caption{Experimental parameters.}
    \label{tab:expparameters}
    \begin{tabular}{|c|c|}
         \hline
         \textbf{Parameter} & \textbf{Value} \\
         \hline
         Sample temperature, T & 300 K \\
         Pump power, $P_\text{\,in}$ & 1 mW \\
         Taper efficiency, $\eta_\text{taper}$ & 0.5 \\
         Taper coupling efficiency $\kappa_\text{e}/\kappa$ & 0.5 \\
         Filtering efficiency, $\eta_\text{filter}$ & 0.15 \\
         SPAD quantum efficiency, $\eta_\text{det}$ & 0.125 \\
         SPAD gate rate, $R_\text{gate}$ & 50 kHz \\
         SPAD effective gate length, $T_\text{gate}$ & 8 ns \\
         SPAD count rate, $R_\text{count}$ & $\sim 500$ s$^{-1}$ \\
         SPAD dark count rate, $R_\text{dark}$ & $\sim 3$ s$^{-1}$ \\
         Balanced detector bandwidth, $B$ & $\sim$ 400 MHz \\
         Heterodyne frequency, $\omega_\text{het}/2\pi$ & 150 MHz \\
         Oscilloscope analog bandwidth & 1 GHz \\
         Oscilloscope sampling rate & 10 GSa/s \\
         \hline
    \end{tabular}
\end{table} \,

\end{document}